\documentclass[a4paper,fleqn,usenatbib]{mnras}

\usepackage{newtxtext,newtxmath}
\usepackage[T1]{fontenc}
\usepackage{ae,aecompl}

\usepackage{graphicx}	% Including figure files
\usepackage{amsmath}	% Advanced maths commands
\usepackage{pdflscape}	% table landscape
\usepackage{multicol}
\usepackage{afterpage}
\usepackage{color}
\usepackage{etoolbox}
\usepackage{hyperref}
\usepackage{booktabs}
\usepackage[utf8]{inputenc}
\usepackage{ulem}

\definecolor{darkmagenta}{rgb}{0.5, 0, 0.5}
\definecolor{darkgreen}{rgb}{0, 0.6, 0.05}
\definecolor{darkred}{rgb}{0.86,0.078,0.235}

\title[MACSJ0416]{BUFFALO Wild Wings: A High Precision Free-Form Lens Model of MACSJ0416 with Constraints on Dark Matter from Substructure and Highly Magnified Arcs}

\author[Derek Perera et al.]{Derek Perera$^{1}$\thanks{E-mail: perer030@umn.edu},
Liliya L. R. Williams$^{1}$, Jori Liesenborgs$^{2}$, Patrick L. Kelly$^{1}$, Sarah H. Taft$^{1}$,
\newauthor{Sung Kei Li$^{3}$, Mathilde Jauzac$^{4,5}$, Jose M. Diego$^{6}$, Priyamvada Natarajan$^{7,8,9}$, Charles L. Steinhardt$^{10}$,}
\newauthor{Andreas L. Faisst$^{11}$, R. Michael Rich$^{12}$, Marceau Limousin$^{13}$}
\\
$^{1}$School of Physics and Astronomy, University of Minnesota, Minneapolis, MN, 55455, USA.\\
$^{2}$UHasselt – Flanders Make, Expertisecentrum voor Digitale Media, Wetenschapspark 2, B-3590, Diepenbeek, Belgium.\\
$^{3}$Department of Physics, The University of Hong Kong, Pokfulam Road, Hong Kong.\\
$^{4}$Department of Physics, Centre for Extragalactic Astronomy, Durham University, South Rd, Durham, DH1 3LE, UK.\\
$^{5}$Department of Physics, Institute for Computational Cosmology, Durham University, South Road, Durham DH1 3LE, UK.\\
$^{6}$Instituto de Fisica de Cantabria (CSIC-UC), Avda. Los Castros s/n, E-39005 Santander, Spain.\\
$^{7}$Department of Astronomy, Yale University, New Haven, CT 06511, USA.\\
$^{8}$Department of Physics, Yale University, New Haven, CT 06520, USA.\\
$^{9}$Black Hole Initiative, Harvard University, 20 Garden Street, Cambridge, MA 02138, USA.\\
$^{10}$Niels Bohr Institute, University of Copenhagen, Lyngbyvej 2, København DK-Ø2100, Denmark.\\
$^{11}$Caltech/IPAC, 1200 E. California Blvd. Pasadena, CA 91125, USA.\\
$^{12}$Department of Physics and Astronomy, University of California, Los Angeles, CA 90095-1547, USA.\\
$^{13}$Aix Marseille Univ, CNRS, CNES, LAM, Marseille, France.
\\
}
\date{Accepted XXX. Received YYY; in original form ZZZ}

\pubyear{2024}

\begin{document}
\label{firstpage}
\pagerange{\pageref{firstpage}--\pageref{lastpage}}
\maketitle

\begin{abstract}
    We present new free-form and hybrid mass reconstructions of the galaxy cluster lens MACS J0416.1$-$2403 at $z=0.396$ using the lens inversion method {\tt GRALE}. The reconstructions use 237 spectroscopically confirmed multiple images from \cite{bergamini23} as the main input. Our primary model reconstructs images to a positional accuracy of 0.191", thus representing one of the most precise reconstructions of this lens to date. Our models find broad agreement with previous reconstructions, and identify two $\sim 10^{12} M_{\odot}$ light-unaffiliated substructures. We focus on two highly magnified arcs: Spock and Mothra. Our model features a unique critical curve structure around the Spock arc with 2 crossings. This structure enables sufficient magnification across this arc to potentially explain the large number of transients as microlensing events of supergiant stars. Additionally, we develop a model of the millilens substructure expected to be magnifying Mothra, which may be a binary pair of supergiants with $\mu \sim 6000$. This model accounts for flexibility in the millilens position while preserving the observed flux and minimizing image position displacements along the Mothra arc. We constrain the millilens mass and core radius to $\lesssim 10^6 M_{\odot}$ and $\lesssim 17$ pc, respectively, which would render it one of the smallest and most compact substructures constrained by lensing. If the millilens is dominated by wave dark matter, the axion mass is constrained to be $\lesssim 3.0 \times 10^{-21}$ eV. Further monitoring of this lens with JWST will uncover more transients, permitting tighter constraints on the structure surrounding these two arcs.
\end{abstract}

\begin{keywords}
galaxies: clusters: individual: MACS J0416.1$-$2403 -- gravitational lensing: strong -- dark matter
\end{keywords}

%%%%%%%%%%%%%%%%%%%%%%%%%%%%%%%%%%%%%%%%%%%%%%%%%%

%%%%%%%%%%%%%%%%% BODY OF PAPER %%%%%%%%%%%%%%%%%%
\section{Introduction}

The mass distributions of galaxy clusters are dominated almost completely by dark matter, with dark matter contributing $\sim100\times$ more to the total mass than baryons \citep{kravstov12}. Observations of individual cluster members and the hot intracluster plasma are, therefore, often insufficient to reconstruct the mass distributions of galaxy clusters. Additional assumptions about the fidelity with which they trace the gravitational potential are required. Strong gravitational lensing of background source galaxies by the cluster offers a powerful technique to accurately reconstruct such mass distributions, and consequently allow for constraints on the nature of dark matter \citep{natarajan24}. This is a direct result of the fact that source galaxies' position in the lens plane and magnification can be solved for with the total mass distribution. Thus, the modeling procedure involves a lens ``inversion'' for this problem, where a suitable mass distribution that can adequately reconstruct multiple image positions is generated to probe the galaxy cluster.

A famous example of a galaxy cluster lens is MACS J0416.1$-$2403 (often shortened to MACSJ0416) detected at $z=0.396$. MACSJ0416 was discovered in the Massive Cluster Survey \citep{ebeling01} and has been extensively studied with numerous Hubble Space Telescope (HST) programs including, the Cluster Lensing And Supernova survey with Hubble \citep[CLASH, ][]{postman12}; the Hubble Frontier Fields \citep[HFF, ][]{lotz17}; the Beyond Ultra-deep Frontier Fields And Legacy Observations \citep[BUFFALO, ][]{steinhardt20}; and Flashlights \citep{kelly22}. Most recently, the James Webb Space Telescope (JWST) began observations of MACSJ0416 with the Prime Extragalactic Areas for Reionization and Lensing Science \citep[PEARLS, ][]{windhorst23} program, with 4 epochs spanning 126 days completed in Cycle 1 \citep{yan23}. These extensive observations have made it the cluster lens with the largest number of multiple images ever discovered to date, with 343 multiple images (237 with spectroscopically confirmed redshifts), permitting unprecedentedly accurate lens modelling and reconstruction of its mass distribution (see Table \ref{tab:summary} for a list of past mass reconstructions of the lens MACSJ0416).

In addition to its many image constraints, MACSJ0416 is of considerable interest due to its elongated bimodal mass structure that is a prototypical feature of actively merging clusters \citep{zitrin13,jauzac14,jauzac15,balestra16}. The merging state of MACSJ0416 implies that it is likely dynamically complex with abundant substructures on varying length scales \citep{jauzac18,cerini23}. Therefore, these properties allow for the results of precise lens models of galaxy clusters informed by large number of images to place constraints on the nature of dark matter, through its potential interaction cross section \citep{kneib11,peter13} or substructure mass fraction \citep{natarajan17,oriordan23,lagattuta23}. Past lens models of MACSJ0416 have constrained the dark matter halo contribution to the total mass to $\sim90\%$ \citep{caminha17,bonamigo17,bonamigo18}.

Meanwhile, the recent discovery of several highly magnified transient stars in the arcs of MACSJ0416 \citep{rodney18,chen19,kaurov19,kelly22,yan23,diego23b} has opened up a new astrophysics research frontier of studying lensed stars at $z \gtrsim 1$. To permit adequate study of such stars, advances in gravitational lensing theory to help model these high magnification ($\mu \gtrsim 1000$) events are currently being developed \citep{venumadhav17,dai21a,meena22}. Probing the mass structures near the critical curves of clusters can increase the resolution of the recovered cluster mass distribution, potentially uncovering individual intracluster stars \citep{Kelly2018}, and dark matter subhalos that do not appear to be associated with visible structures \citep{williams24}. The time domain nature of these lensed transients has been used to constrain the probability of microlensing \citep{dai21b,li24}; stellar abundance at high redshifts \citep{diego24}; and the properties of dark matter \citep{diego18,oguri18,dai20b}. Similarly, accurate modelling of the regions near critical curves are crucial to help elucidate properties of high redshift stellar systems \citep{pascale23,claeyssens23,klein24}.

Table \ref{tab:summary} summarizes many of the recent lens mass reconstructions for MACSJ0416 and their results. The identification of increasing numbers of multiple images has contributed to increased precision in the models regardless of adopted methodology. In general, parametric lens models account for cluster member galaxies and the cluster dark matter halo with analytic density profiles such as Navarro-Frenk-White \citep[NFW, ][]{nfw97}; pseudo-isothermal \citep{Natarajan1997} or pseudo-Jaffe \citep{keeton01}. Parametric models have the advantage of being physically motivated by properties of the cluster and being directly comparable to cosmological simulations, although this can lead to bias and the inability to identify smaller scale features of the mass distribution and to recover small-scale substructures that may not be associated with light. An alternative approach is provided by free-form models, which do not include cluster light information as a prior. These are advantageous in their flexibility, offering an unbiased view of the lens, but may predict properties of the mass distribution that are physically disfavored. Hybrid models offer a middle ground approach, incorporating physically motivated parametric priors on top of a free-form lens framework. In this work, we model MACSJ0416 using the free-form and hybrid methods with {\tt GRALE}, a lens inversion technique making use of a genetic algorithm \citep{liesenborgs06}. This is the second model of MACSJ0416 using {\tt GRALE}, after the work by \cite{sebesta16}. Our model presented is here is the most precise to date for MACSJ0416, offering new constraints on substructure and dark matter.

With our new model, we also study the structure surrounding two highly magnified arcs\footnote{These two arcs are the origin of the titular ``wings'', as they look like wings straddling the body of the cluster.}, Spock ($z = 1.005$) and Mothra ($z = 2.091$). The Spock arc has been of recent interest due to its complex local mass structure and the discovery of numerous transients across the arc \citep{rodney18,kelly22,yan23}. Various interpretations ranging from recurrent novae \citep{rodney18} to microlensing by intra-cluster stars \citep{diego24} have been suggested to explain the transients, with more preference for the latter explanation with the increased observational cadence. An accurate model of this region will provide tight constraints on the abundance of lensed supergiants in the source \citep{diego24} and the frequency of microlensing. At this point, no lens models in the literature have been able to successfully replicate all the observations of the Spock arc. Observations of the Mothra transient in the Mothra arc \citep{diego23b,yan23} have found that it has been visible for longer than 8 years without a confident counterimage. This suggests that Mothra is not a microlensing event, but rather a millilensing event by a $\gtrsim 10^4$ $M_{\odot}$ mass substructure. The exact size of this millilens, however, is unconstrained at present and could range from $\sim 10^6 M_{\odot}$ \citep{diego23b} to $\sim 10^9 M_{\odot}$ \citep{abe23}. Therefore, sophisticated lens models of this millilens have the potential to constrain it to be the smallest substructure found with lensing, and can place tight constraints on dark matter \citep{diego23b}.

In this work, we study a variety of free-form and hybrid lens models of MACSJ0416. Overall, all of our models reconstruct the observed positions of the multiple images to high precision, and predicts the existence of two dark substructures that appear to be 
unassociated with light, with their reality needing more scrutiny. Our main free-form lens model reconstructs a multiple critical curve crossing structure for the Spock arc that can adequately explain its observed high magnification and the transient detection rate, and is one of the few lens models capable of doing so. Because of this, we intensively scrutinize this result to ensure that it is robust. We also test the interpretations of the Spock transients as stellar variability or microlensing by including them as explicit sources in separate lens models. The second half of this work presents a millilens modelling method that we use to constrain the mass and core radius of the millilens magnifying the Mothra transient. We use these constraints alongside the mass substructures identified in the model to place constraints on different dark matter models.

In Section \ref{txt:data}, we present the image and cluster galaxy datasets that we use in our modelling. Section \ref{txt:recon} describes our lens reconstruction method with {\tt GRALE} and a discussion of all the free-form and hybrid models we generate. Section \ref{txt:results} presents our results, including those for the Spock and Mothra arcs. Section \ref{txt:conclusions} discusses the implications of our results and avenues for future study of MACSJ0416. For this work, we assume a flat $\Lambda$CDM cosmology with $\Omega_M = 0.27$, $\Omega_{\Lambda} = 0.73$, and $H_0 = 70$ km s$^{-1}$ Mpc$^{-1}$. At the lens redshift $z_d = 0.396$, 1 arcsec corresponds to 5.386 kpc.

\begin{table*}
	\caption{Past Lens Reconstructions of MACS J0416.1-2403}
	\begin{tabular}{ccccccc} 
		\hline
		Lens Model & Method & $N_{\rm im}$ ($N_{\rm im,z}$) & Degrees of Freedom & $\Delta_{RMS}$ & Spock Arc CC Crossings & Possible Substructures  \\
		\hline
		\cite{johnson14} & {\tt LensTool} (Par) & 50 (26) & 21 & 0.51" & N/A & N/A \\
		\cite{jauzac14} & {\tt LensTool} (Par) & 149 (26) & N/A & 0.68" & N/A & 2 \\
    	\cite{sebesta16} & {\tt GRALE} (FF)  & 149 (26) & N/A & N/A & N/A & N/A \\
		\cite{bergamini19} & {\tt LensTool} (Par) & 102 (102) & 110 & 0.61" & N/A & 0 \\
		\cite{gonzalez20} & {\tt LensTool} (Par) & $\sim$171 (N/A) & N/A & N/A & N/A & 5 \\
  		\cite{raney20} & {\tt Keeton} (Par) & 95 (95) & N/A & 0.52" & 2 & N/A \\
		\cite{bergamini21} & {\tt LensTool} (Par) & 182 (182) & 202 & 0.40" & 1 & 0 \\
		\cite{richard21} & {\tt LensTool} (Par) & 198 (198) & N/A & 0.58" & N/A & N/A \\
		\cite{limousin22} & {\tt LensTool} (Par) & 182 (182) & N/A & 0.62" & N/A & 0 \\ 
  		\cite{bergamini23} & {\tt LensTool} (Par) & 237 (237) & 268 & 0.43" & 1 & 0 \\
		\cite{cha23} & {\tt MARS} (FF) & 236 (236) & N/A & 0.0836" & N/A & 0 \\
		\cite{diego23a} & {\tt WSLAP+} (H) & 343 (237) & N/A & N/A & 1 & 0 \\
  		\cite{diego24} & {\tt WSLAP+} (H) & 214 (214) & N/A & N/A & 1 & 0 \\
  		\cite{rihtarsic24} & {\tt LensTool} (Par) & 303 (303) & 354 & 0.53" & N/A & N/A \\
		This Work & {\tt GRALE} (FF) & 237 (237) & N/A & 0.191" & 2 & 2 \\
		
		\hline
	\end{tabular}\\
\medskip{A summary table of recent lens models for MACS J0416.1-2403. \cite{remolina18} provides a review of lens models prior to $\sim$2018. The columns list the following: lens model reference, the reconstruction method and type (``Par'' for parametric, ``H'' for hybrid, and ``FF'' for free-form), the number of images $N_{\rm im}$ used (Number of images with spectroscopic redshifts $N_{\rm im,z}$), model degrees of freedom (where reported), the lens plane RMS $\Delta_{RMS}$, the number of critical curve crossings present in the Spock Arc, and the number of identified substructures. Anywhere listed ``N/A'' indicates that the study did not report the information.}
\label{tab:summary}
\end{table*}

\section{Data}\label{txt:data}

\begin{figure*}
\includegraphics[trim={4.5cm 0.3cm 4.5cm 0cm},clip,width=0.75\textwidth]{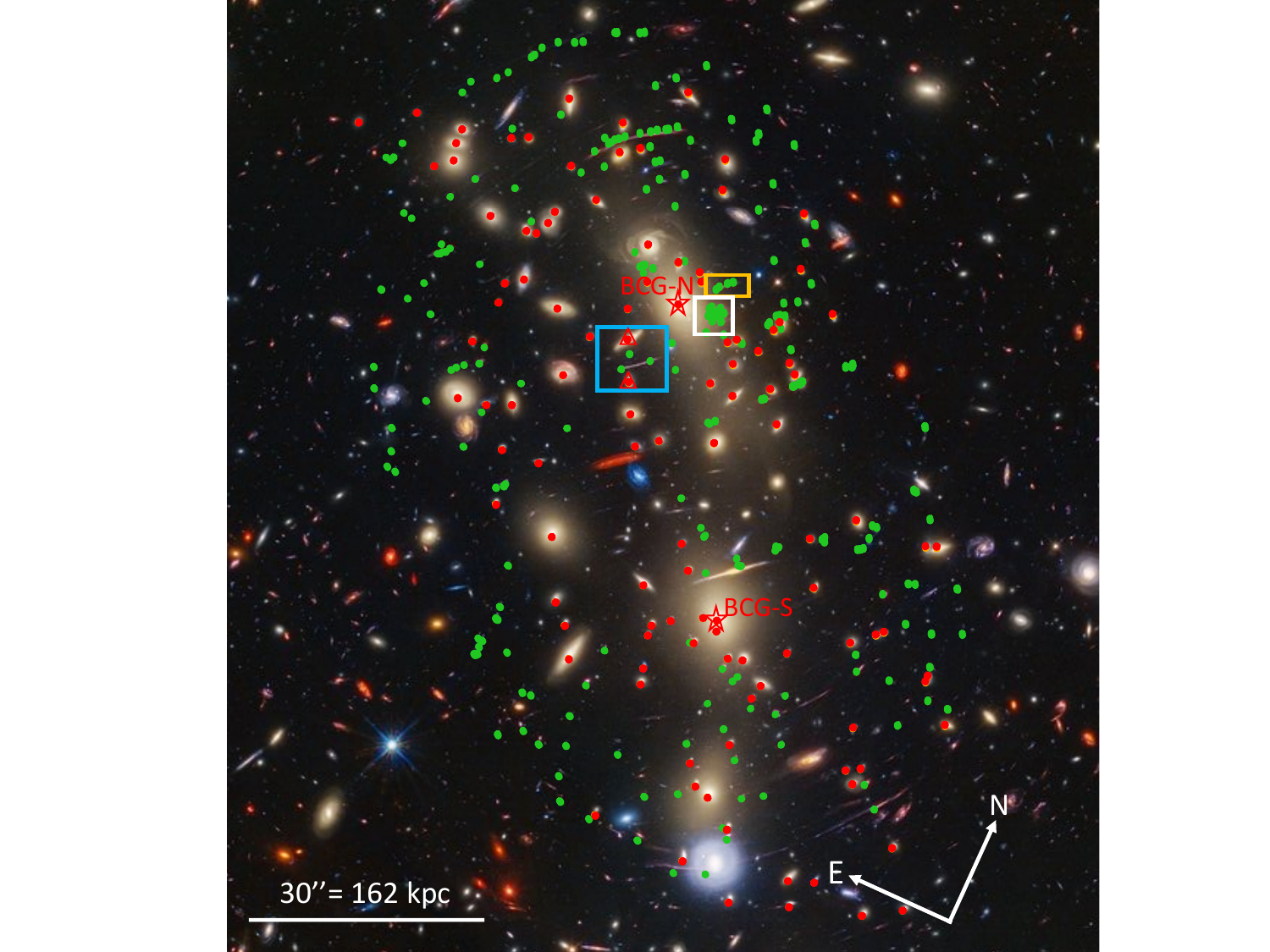}%{Figures/MACSJ0416image.png}
\caption{Panchromatic image of MACSJ0416 (credits to \href{https://webbtelescope.org/contents/media/images/2023/146/01HDHATAQXM532HCNQN6BQ79BC?news=true}{NASA, ESA, CSA, and STScI}). Green dots refer to observed multiple images and red dots refer to cluster member galaxies. The northern and southern BCGs (BCG-N and BCG-S) are explicitly highlighted with red stars, while the Spock North and South galaxies (Spock-N and Spock-S) are highlighted with red triangles. The blue, white, and orange squares enclose the Spock, Warhol, and Mothra arcs, respectively, which we study in detail in this work. These arcs form the titular ``wings'' of the cluster with their location on either side of central BCG.}
\label{fig:bigmosaic}
\end{figure*}

%filter central wavelengths are: F435W at 432.9868nm, F606W at 592.1892nm, and F814W at 804.5533nm

Figure \ref{fig:bigmosaic} shows a color composite HST image of MACSJ0416.1$-$2403. We make use of the catalog of 237 spectroscopically identified multiple images from 88 distinct background sources as compiled by \cite{bergamini23}\footnote{The dataset is available in the article as well as at the following page: \url{https://cdsarc.cds.unistra.fr/viz-bin/cat/J/A+A/674/A79}}. Their image catalog builds off the prior analysis in \cite{bergamini21} using Hubble multi-color imaging and VLT Multi-Unit Spectroscopic Explorer (MUSE) spectroscopy \citep{richard21}. \cite{bergamini21} identified 182 multiple images which increased to the current 237 \citep{bergamini23} with the inclusion of primarily bright knots in extended sources. 

For this catalog, Hubble imaging was performed in 7 filters with the HFF program \citep{lotz17} and in 16 filters with the CLASH survey \citep{postman12,balestra16}. Deep VLT MUSE spectroscopy was obtained with a total integration time of 17.1h \citep{vanzella21}. We refer to \cite{bergamini21} for a complete discussion of the observations. The resulting image catalog represents the largest dataset of secure multiple image positions with spectroscopic redshifts for any lens system. This makes it an ideal image dataset for use with {\tt GRALE}. We note that, recently, the PEARLS project \citep{windhorst23} observed MACSJ0416, thereby increasing the total number of identified multiple images to 343 \citep{diego23a}. However, since many of these new images have yet to be spectroscopically confirmed, we do not include them in this work. We restrict our analysis to multiple images with spectroscopic redshifts because {\tt GRALE} has previously been shown to have reduced scatter between observed and reconstructed images, when using spectroscopically confirmed images only \citep{johnson14,grillo15,remolina18}. 

As we discuss further in Section \ref{txt:input}, not included in the catalog of 237 images are the numerous transients that have been discovered recently with the Flashlights and PEARLS programs. In total, 19 transients (thought to be a result of microlensing near the cluster critical curve) have been discovered in MACSJ0416, with the vast majority in the Spock ($z_s = 1.005$) and Warhol ($z_s = 0.94$) arcs \citep{rodney18,chen19,kaurov19,kelly22,yan23}. In general, these transients are discovered due to temporary increase in brightness, aided most likely by microlensing, and captured by time domain observations of $\sim$2-6 epochs (on the order of 100s of days) at a depth of $m_{AB} \sim 29$. One persistent high magnification event, Mothra, has been visible for longer than 8 years, suggesting a more permanent substructure causing its high magnification \citep{diego23b} rather than a transient event. 

In this work, we study Mothra and various transients in the Spock arc. Observations of Mothra are presented in \cite{yan23} with 8 JWST filters across 4 epochs spanning 126.1 days. SED fitting of Mothra identify the source as a system of 2 binary supergiant stars as discussed in \cite{diego23b}. Observations of Spock are plentiful, but we make use of the observed transients S1/S2 \citep{rodney18}, F1/F2 \citep{kelly22}, and D21-S1/S2 \citep{yan23}. We discuss respective interpretations of these transients in Section \ref{txt:input}. We include separate lens models making use of these transients as specific constraints, those results are shown in Table \ref{tab:models}.

Finally, cluster member galaxies are identified in the catalog from \cite{tortorelli23}\footnote{The dataset is available at the following page: \url{https://cdsarc.cds.unistra.fr/viz-bin/cat/J/A+A/671/L9}}. Properties of cluster member galaxies are measured from the HFF data in 3 Advanced Camera for Surveys (ACS) optical filters and 4 Wide Field Camera 3 (WFC3) NIR filters. The structural parameters of these galaxies are utilized as priors in our hybrid lens models. We note that in the Southern region of the cluster there exists a bright foreground galaxy at $z = 0.112$. We do not include this in any of our models as its mass contribution is relatively small and it is established as not being a cluster member.

\section{Lens Reconstructions}\label{txt:recon}

\subsection{Lens Reconstruction with Grale}\label{txt:grale}
For this paper we use the free-form lens reconstruction code {\tt GRALE} \footnote{{\tt GRALE} is publicly available, and the software and tutorials can be found at the following page: \url{https://research.edm.uhasselt.be/~jori/grale2/index.html}} to perform lens inversions. {\tt GRALE} utilizes a flexible inversion method based on a genetic algorithm that optimizes a mass basis on an adaptive grid \citep{liesenborgs06,liesenborgs07,liesenborgs20}. For our reconstructions, we utilize a mass basis of projected Plummer spheres, which have projected surface mass densities of:
\begin{equation}\label{eq:plummerdens}
    \Sigma(\boldsymbol{\theta}) = \frac{M}{\pi D_{\mathrm{d}}^2}\frac{\theta_\mathrm{P}^2}{(\boldsymbol{\theta}^2 + \theta_\mathrm{P}^2)^2},
\end{equation}
and lens potentials of:
\begin{equation}\label{eq:plummerlpot}
    \psi(\boldsymbol{\theta}) = \frac{2GM D_\mathrm{ds}}{c^2 D_\mathrm{s} D_\mathrm{d}} \ln{(\boldsymbol{\theta}^2 + \theta_\mathrm{P}^2)},
\end{equation}
where $\theta_\mathrm{P}$ is the characteristic angular width of the Plummer sphere, $M$ is its total mass, and $D$'s are angular diameter distances between the observer, {\it s}ource and {\it d}eflector. {\tt GRALE} uses a genetic algorithm to optimize the respective weights of each Plummer in the grid based primarily on how well images backprojected into the source plane overlap with one another. These weights can be determined by several fitness measures, of which we use two (in order of priority):
\begin{itemize}
    \item ``pointimagenull'': Regions in the lens plane where no images form are subdivided independently into a grid of triangles. This grid of triangles is backprojected into the source plane where triangles overlapping with estimated sources are penalized. This amounts to disfavouring maps that produce extraneous images at each generation. For a full description of this criterion, see \cite{zitrin10}.
    \item ``pointimageoverlap'': Images of the same source are backprojected into the source plane. If the images overlap more, then the map has a better fitness. For point images, this amounts to the source plane distance between backprojected images. Importantly, the scale of this overlap is determined by the region defined by all the backprojected images. This is designed to defend against overfocusing. For more information, see \cite{zitrin10}.
\end{itemize}
Once optimized, the grid for the basis functions is subdivided further, with regions of greater density refined more significantly. This process continues for many iterations. The best lens model is the subdivision grid with the best overall fitness values. The number of subdivisions, and therefore the number of Plummers, can vary for different lens models (typically $\sim$1000-5000 Plummers per model). The resulting grid of diversely sized Plummers can then be used to calculate $\Sigma(\theta)$ (Eq. \ref{eq:plummerdens}) and $\psi(\theta)$ (Eq. \ref{eq:plummerlpot}) at any position in the lens plane to arbitrary precision given their analytic functions. In the end, a single {\tt GRALE} run consists of a subdivision grid of thousands of Plummers with unique sizes and weights as determined by the optimized fitness criteria.

The default mass grid for {\tt GRALE} is a grid of Plummers, which we adopt as the baseline grid for all lens models we generate. It is possible to add parametric lens models\footnote{Lens models built into {\tt GRALE} can be found at the following page: \url{https://research.edm.uhasselt.be/~jori/grale2/grale_lenses.html}} on top of the grid of Plummers. Instead of a free-form lens model, this would instead be a hybrid lens model since parametric lens components would be added to a free-form skeleton model. In this case, the same reconstruction procedure as described here would follow, with the basis grid consisting of the Plummers and chosen parametric models being optimized with the same fitness criteria \citep{liesenborgs20}. It is important to note that only the weights of these parametric components is optimized in this hybrid approach.

Since each {\tt GRALE} run will produce a slightly different lens reconstruction, we take our lens models to be the average of 40 {\tt GRALE} runs. The decision to average over 40 runs is motivated by limitations of computational resources and is consistent with previous reconstructions using {\tt GRALE} \citep{williams19,sebesta19,gho20,ghosh21,ghosh23,perera24}. Furthermore, averaging over many runs defends against degenerate mass features and allows for quantification of uncertainties in the lens model.

Lastly, it is important to consider the fact that this mean lens reconstruction need not be the best lens reconstruction. Backprojected images for the mean lens model may not converge to a well-defined source position. To account for this, we use the source position optimization method from \cite{perera24}. This optimization does not alter the mass distributions produced by the genetic algorithm, as it is done after all {\tt GRALE} runs are completed. For a full discussion of this procedure, we direct the reader to section 3.2 of \cite{perera24}. As a brief summary of the method, we use a Metropolis-Hastings algorithm to optimize the source position to better fit the observed images. This amounts to minimizing the following likelihood:
\begin{equation}\label{eq:galposterior}
   \ln \left(P(\boldsymbol{\beta})\right) = -\frac{1}{2}\sum_\mathrm{i}\left[\left(\frac{x'_\mathrm{i} - x_\mathrm{i}}{\sigma_\mathrm{x}}\right)^2 + \left(\frac{y'_\mathrm{i} - y_\mathrm{i}}{\sigma_\mathrm{y}}\right)^2 \right], 
\end{equation}
where ($x_i$,$y_i$) are the observed image positions, ($x'_i$,$y'_i$) are the reconstructed image positions at the sampled source position $\boldsymbol{\beta}$, and ($\sigma_x$, $\sigma_y$) are the image position uncertainties defined to be 0.04", corresponding to the astrometric precision of HST. We note that 14 of the 88 sources do not have detected HST counterparts \citep{caminha17}. These sources are detected primarily with Lyman-$\alpha$ emission using MUSE Wide Field Mode, which would imply a positional uncertainty of 0.2" for these images, larger than our assumed $\sigma_\mathrm{x}$ and $\sigma_\mathrm{y}$. However, recent JWST observations of MACSJ0416 have confirmed the image positions that we use here \citep{diego23a}, justifying our use of smaller uncertainties. Extended morphology of the sources is also not applicable to our procedure since these sources are included as multiply imaged point sources of knots within the extended arcs, which has been shown to be adequate in constraining critical curve locations \citep{bergamini21}. 

We impose the same flat prior as \cite{perera24} on each sampled $\beta$ such that it is uniformly sampled in the region defined by minimum and maximum source positions of the 40 original runs. The result of this procedure is what we use as the complete lens model, equivalent to the averaged lens model of 40 runs with optimized source positions.

\subsection{Model Inputs}\label{txt:input}

Since the default settings of {\tt GRALE} only require observed multiple image data and redshifts as input, it is important that the data be of the highest possible quality. MACSJ0416 is therefore an ideal candidate for a lens reconstruction at the present time since it has 237 spectroscopically confirmed images, making this the largest sample of gravitationally lensed sources. Our main lens model, generated with the process described in Section \ref{txt:grale}, uses all 237 of these images as input. We refer to this lens model FF00.

Not included in the aforementioned 237 images as noted previously are the plethora of transient events that have been discovered in MACSJ0416. The primary reason for this is that these objects typically lack counterimages, and therefore it is difficult to ascertain their source positions. In fact, many of these transients are thought to be influenced primarily by microlensing or millilensing. However, some notable exceptions are the transients discovered in the Spock arc region, where it has been hypothesized that some transients are in fact counterimages of one another, implying that they originate from the same source. Specifically, in \cite{rodney18} two fast transients (S1 and S2) were discovered in the Spock arc and postulated to be from the same region of the source galaxy but distinct events in time. Based on the lens model used, the transients can be explained by various phenomena ranging from independent eruptions from the surface of a luminous blue variable (LBV) star to separate stellar microlensing events \citep{rodney18}. Recent observations from HST's Flashlights and JWST's PEARLS programs have discovered many more transient events in the Spock Arc. \cite{kelly22} finds two distinct transients (F1 and F2) nearby to the lens plane positions of S1 and S2 and hypothesize that these could be counterimages of S1 and S2. It is also entirely possible that F1 and F2 are instead counterimages of one another. \cite{yan23} reports the discovery of 4 more transients, of which 3 (D21-S1, D21-S2, and D31-S4) have secure photometry, consistent with expectations of $\sim 1-5$ transients per pointing in the Spock arc \citep{diego24}. Of the 3 transients, D21-S1 and D21-S2 are discovered in the same epoch, indicating that they may be counterimages. All these mentioned transients are shown in Figure \ref{fig:spockCC}.

With all these transients in the Spock Arc, studying the density profile and critical curve structure of the local region becomes extremely complicated. Since some of the transients are potentially counterimages of one another, we create 4 additional free-form models to test these hypotheses (we adopt the notation where ``A/B'' indicates that transients A and B are counterimages):
\begin{itemize}
    \item FF11: Input main 237 images along with S1/F1 and S2/F2 as explicit counterimages of the same source. This tests the scenario postulated by \cite{kelly22}.
    \item FF12: Input main 237 images along with S1/S2 and F1/F2 as explicit counterimages of the same source. This tests an alternative scenario to FF11.
    \item FF11+D: Same as FF11 but also including D21-S1 and D21-S2 as explicit counterimages of the same source.
    \item FF12+D: Same as FF12 but also including D21-S1 and D21-S2 as explicit counterimages of the same source.
\end{itemize}
Comparing each of these free-form models with FF00 will help determine if any of the transients could in fact be counterimages of one another. 

In addition to free-form models, {\tt GRALE} has the capability to do hybrid lens inversions as briefly described in Section \ref{txt:grale}. For these we include parametric models for the brightest cluster galaxies (BCGs) and the northern and southern cluster member galaxies surrounding the Spock arc, Spock-N and Spock-S, respectively (see Figure \ref{fig:spockCC} for the identification of these 2 galaxies). To compare with FF00, we further present two hybrid lens models using all 237 images as input: 
\begin{itemize}
    \item H-Ser: includes BCG-N, BCG-S, Spock-N, and Spock-S as circular Sersic models. Equation \ref{eq:sersic} gives the density profile for a Sersic model.
    \item H-NFW: includes BCG-N, BCG-S, Spock-N, and Spock-S as NFW models. Equation \ref{eq:nfw} gives the density profile for an NFW model.
\end{itemize}
In both cases, the inclusion of explicit parametric models for Spock-N and Spock-S is motivated by the interest in understanding the critical curve structure in the Spock Arc region. 

Since {\tt GRALE} optimizes the weights of all components in the complete mass basis (which will include the aforementioned parametric components), some of the input parameters for the NFW and Sersic components only need to be approximately representative of each galaxy. Therefore, to initialize the parametric components of both models, we use the observed measurements of the structure and brightness of each respective galaxy as presented in \cite{tortorelli23}. These yield the input parameters for both models as presented in Tables \ref{tab:hybridS} and \ref{tab:hybridN} for H-Ser and H-NFW, respectively. See Section \ref{txt:sersic} and \ref{txt:nfw} for a full derivation and discussion of how we attained the input parameters for H-Ser and H-NFW, respectively. H-Ser and H-NFW only include the aforementioned cluster galaxies, with the remaining cluster member galaxies excluded for this analysis. This choice potentially biases the resultant model, as excluded mass contributions from these cluster member galaxies may contribute to different predictions of magnification and shear. However, since we are primarily interested in the reconstruction of the Spock arc, only Spock-N and Spock-S are critical to the model, and the remaining cluster member galaxies will have negligible effect. Furthermore, since the cluster member mass contribution is suppressed $\gtrsim 10"$ away from BCG-N \citep{bonamigo18}, we do not expect this bias to be significant for our purposes. We note that the best way to examine this bias is to build a full hybird model using {\tt GRALE} including all known cluster member galaxies, which is the subject of a future work.

\begin{table}
    \caption{Hybrid Lens Model Parameters - Sersic (H-Ser)}
    \centering
    \begin{tabular}{ccccc}
    \hline
        Galaxy & $\Sigma_{\rm cen}$ [$M_{\odot}$ kpc$^{-2}$] & $\theta_S$ [pc]  & $M_{\star}$ [$M_{\odot}$] & $R_e$ [kpc] \\
    \hline
        BCG-N & $2.46 \times 10^{10}$ & 2.55 & $2.03 \times 10^{10}$ & 8.81 \\
        BCG-S & $1.50 \times 10^{11}$ & 3.13 & $1.86 \times 10^{11}$ & 10.83\\
        Spock-N & $3.82 \times 10^{11}$ & 0.54 & $1.43 \times 10^{10}$ & 1.88 \\
        Spock-S & $3.25 \times 10^{10}$ & 0.31 & $4.08 \times 10^{8}$ & 1.08 \\
    \hline
    \end{tabular}
    \medskip{Input parameters used in H-Ser and all hybrid models using Sersic models for each galaxy. $\Sigma_{\rm cen}$ and $\theta_S$ are the central surface mass density and angular scale, respectively, directly used to parametrize each Sersic. $M_{\star}$ is the stellar mass of the galaxy estimated using the observed correlation with effective radius $R_e$ \citep{ulgen22}. $R_e$ is measured in HST F160W and presented in \cite{tortorelli23}. See Appendix \ref{txt:sersic} for a full discussion of these model parameters.}
    \label{tab:hybridS}
\end{table}

\begin{table}
    \caption{Hybrid Lens Model Parameters - NFW (H-NFW)}
    \centering
    \begin{tabular}{ccccc}
    \hline
        Galaxy & $\rho_s$ [$M_{\odot}$ kpc$^{-3}$] & $r_s$ [kpc]  & $M_{\rm vir}$ [$M_{\odot}$] & $R_{\rm vir}$ [kpc] \\
    \hline
        BCG-N & $1.92 \times 10^{7}$ & 47.88 & $3.47 \times 10^{13}$ & 383.04 \\
        BCG-S & $9.59 \times 10^{6}$ & 58.86 & $3.21 \times 10^{13}$ & 470.87 \\
        Spock-N & $2.60 \times 10^{8}$ & 8.17 & $2.66 \times 10^{12}$ & 81.74 \\
        Spock-S & $2.97 \times 10^{8}$ & 4.70 & $5.75 \times 10^{11}$ & 46.96 \\
    \hline
    \end{tabular}
    \medskip{Input parameters used in H-NFW and all hybrid models using NFW models for each galaxy. $\rho_s$ and $r_s$ are the scale density and scale radius for the NFW profile, respectively. $M_{\rm vir}$ is the virial mass estimated with the virial radius $R_{\rm vir}$ and velocity dispersion. $R_{\rm vir}$ is estimated using its relation with the observed $R_e$ \citep{huang17}. See Appendix \ref{txt:nfw} for a full discussion of these model parameters.}
    \label{tab:hybridN}
\end{table}

\section{Results}\label{txt:results}

\subsection{Projected Surface Mass Density Distribution}\label{txt:surfdensmaps}

\begin{table*}
	\caption{Lens Reconstructions of MACS J0416.1-2403 in this Work}
	\begin{tabular}{cccccc} 
		\hline
		Lens Model & $\Delta_{RMS}$ & Transient Counterimages & Spock Arc $\langle \Delta \boldsymbol{\theta} \rangle$ & Substructure Mass [$10^{11} M_{\odot}$] & $M(< 200 kpc)$  [$10^{13} M_{\odot}$] \\
            & & & & (M1,M2) & (BCG-N,BCG-S) \\
		\hline
  		FF00 & 0.191" & None & 0.111" & ($9.5 \pm 0.5$, $5.7 \pm 0.2$) & ($14.34 \pm 0.02 $, $14.87 \pm 0.02$) \\
  		FF11 & 0.204" & S1/F1,S2/F2 & 0.145" & ($10 \pm 0.4$, $5.6 \pm 0.2$) & ($14.40 \pm 0.02 $, $14.92 \pm 0.02$) \\
  		FF12 & 0.213" & S1/S2,F1/F2 & 0.160" & ($9.7 \pm 0.4$, $6.0 \pm 0.3$) & ($14.39 \pm 0.02 $, $14.91 \pm 0.02$) \\
  		FF11+D & 0.209" & S1/F1,S2/F2,D21-S1/S2 & 0.413"  & ($11 \pm 0.5$, $7.4 \pm 0.3$) & ($14.41 \pm 0.02 $, $14.95 \pm 0.02$) \\
  		FF12+D & 0.201" & S1/S2,F1/F2,D21-S1/S2 & 0.251" & ($9.7 \pm 0.4$, $5.2 \pm 0.2$) & ($14.40 \pm 0.02 $, $14.92 \pm 0.02 $) \\
  		H-NFW & 0.206" & None & 0.163" & ($11 \pm 0.5$, $5.0 \pm 0.2$) & ($14.51 \pm 0.02 $, $15.08 \pm 0.02 $) \\
  		H-Ser & 0.207" & None & 0.236" & ($11 \pm 0.5$, $7.3 \pm 0.2$) & ($14.31 \pm 0.02 $, $14.86 \pm 0.02$) \\
		\hline
	\end{tabular}\\
\medskip{A summary table of the lens models generated in this work. FF00 is our main lens model, and all lens models and names are defined in Section \ref{txt:input}. $\Delta_{RMS}$ is the lens plane RMS (see equation \ref{eq:rms}). If the model included any transients as explicit counterimages, these are stated (e.g. S1/F1 in FF11 corresponds to transients S1 and F1 included as explicit counterimages of the same source star). The mean image plane separation $\langle \Delta \boldsymbol{\theta} \rangle$ for the Spock arc images in the model is also given to compare the precision of the model at the Spock Arc. We also give the mass of the two substructures M1 and M2, and the mass within 200 kpc of BCG-N and BCG-S ($M(< 200 kpc)$). $M(< 200 kpc)$ is not background subtracted. }
\label{tab:models}
\end{table*}

Here we describe specific results of our lens models. Table \ref{tab:models} summarizes these main results. Unless otherwise noted, quoted uncertainties are standard deviations of the measured quantities from the sample of 40 {\tt GRALE} runs.

\subsubsection{Free-Form Model: FF00}\label{txt:freeformdensdist}

\begin{figure}
    \centering
    \includegraphics[trim={5.1cm 0.35cm 5.1cm 0.35cm},clip,width=0.49\textwidth]{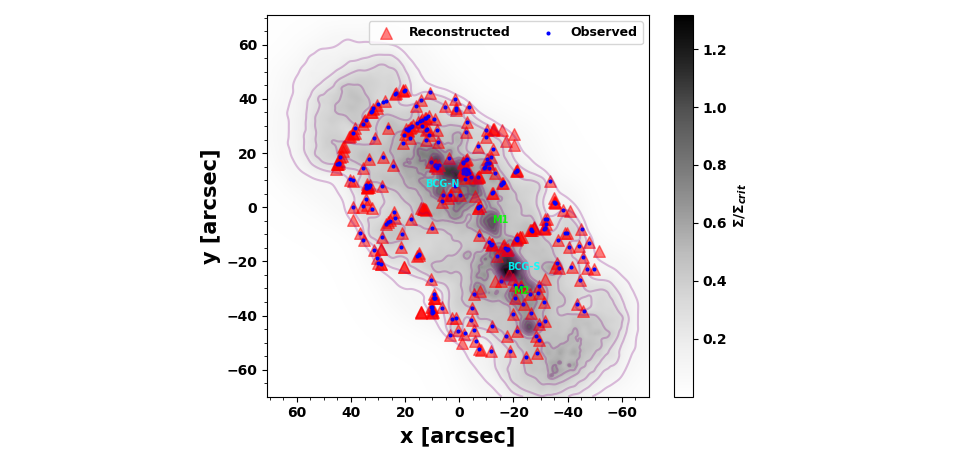}
    \includegraphics[trim={0cm 0cm 0cm 0cm},clip,width=0.49\textwidth]{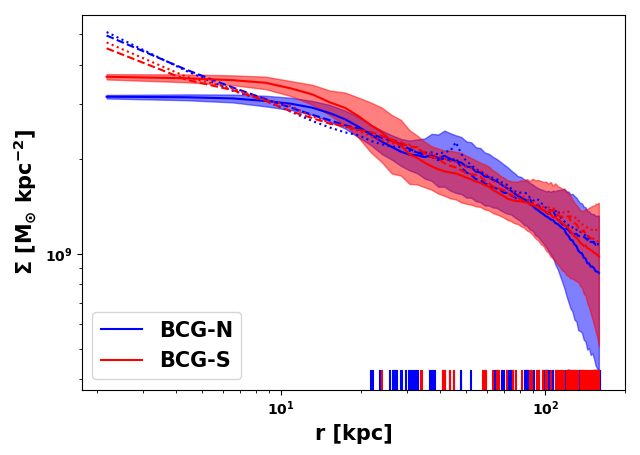}
\caption{{\it Top:} Projected surface mass density distribution for our main lens model FF00. Axes are presented in arcseconds with respect to the zero point defined to be the mean position of all 237 observed images (blue dots). Reconstructed images from the model are shown with red triangles. The two main mass peaks correspond to the BCGs and are labelled accordingly. Two light unaccompanied mass peaks, M1 and M2, are labelled in green. Here, 1 arcsec is equivalent to 5.386 kpc and $\Sigma_{\rm crit} = 8.3985 \times 10^{10} M_{\odot}$ arcsec$^{-2}$ for the nearest source ($z = 0.94$). The separation between BCG-N and BCG-S is $\sim 250$ kpc. The contour lines are separated by $\Delta\kappa_{z=0.94} = 0.1$ in surface mass density. {\it Bottom:} Circularly averaged surface mass density profiles from the BCG-N (blue) and BCG-S (red) for FF00. The shaded regions list the 68\% confidence level for each profile. Vertical dashes indicate the image positions relative to their respectively coloured BCG. For comparison, the circularly averaged surface mass density profiles for \citet{bergamini23} (dashed) and \citet{diego23a} (dotted) are shown with respect to each BCG.} 
\label{fig:massdist}
\end{figure}

\begin{figure}
\includegraphics[trim={0cm 0cm 0cm 0cm},clip,width=0.49\textwidth]{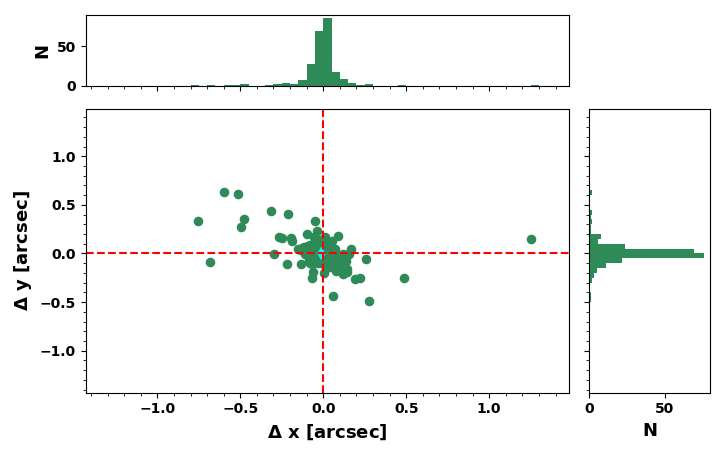}

\caption{Image separations $\boldsymbol{r_{i,\rm obs}} - \boldsymbol{r_{i,\rm rec}}$ along the $x$ and $y$ axes. The histograms show the distributions of these separations on both axes. The red dashed lines indicate a separation of 0". The cyan diamond indicates the mean image separation at (-0.011", 0.004").}
\label{fig:RMS}
\end{figure}

The top panel of Figure \ref{fig:massdist} shows the projected surface mass density distribution for our main lens model FF00. Morphologically, it is similar to previous recent reconstructions of MACSJ0416 \citep{jauzac14,jauzac15,caminha17,bergamini21,bergamini23,cha23,diego23a}. The two main mass peaks in the vicinity of BCG-N and BCG-S are displaced by $\sim$1.5" and $\sim$1.7", respectively. These minor offsets are within the uncertainty for the cluster due to the lack of observational constraints near the centers of the BCGs. The bottom panel of Figure \ref{fig:massdist} presents the circularly averaged density profiles about BCG-N and BCG-S. Both profiles are broadly similar out to $\sim$200 kpc, which is consistent with recent lens models \citep{bergamini23,diego23a,diego24}. One subtle but interesting feature of our model is that BCG-S is more massive than BCG-N within $\sim$30 kpc, which differs from the models from \cite{diego23a,diego24} but is consistent with \cite{bergamini23}. The mass within 200 kpc is $1.434 \pm 0.002 \times 10^{14} M_{\odot}$ and $1.487 \pm 0.002 \times 10^{14} M_{\odot}$ for BCG-N and BCG-S, respectively. This is in reasonable agreement with \cite{diego23a}, who finds 1.72 $\times 10^{14} M_{\odot}$ and 1.77 $\times 10^{14} M_{\odot}$ for BCG-N and BCG-S, respectively.

A common way to quantify the quality of fit of a lens model is with the lens plane root-mean-square (RMS) separation between observed and reconstructed image positions:
\begin{equation}\label{eq:rms}
    \Delta_{\rm RMS} = \sqrt{\frac{\sum_{\rm i}^{N_{\rm im}} |\boldsymbol{r_{\rm i,\rm obs}} - \boldsymbol{r_{\rm i,\rm rec}}|^2}{N_{\rm im}}},
\end{equation}
where $\boldsymbol{r_{\rm i,\rm obs}}$ and $\boldsymbol{r_{\rm i,\rm rec}}$ are the observed and reconstructed $i$th image positions, respectively, and $N_{\rm im}$ is the total number of images. Figure \ref{fig:RMS} shows the image displacement distribution along each axis and total histogram of all image separations. For FF00, $\Delta_{\rm RMS} =$ 0.191", making FF00 more accurate than all parametric models for this cluster and one of the most accurate lens models of MACSJ0416. We note that the RMS separation prior to source position optimization was 0.478", highlighting the effectiveness of this method for reducing image scatter. Additionally, $\sim$86\% of reconstructed images have an image separation lower than our $\Delta_{RMS}$ value. In Figure \ref{fig:RMS}, a weak correlation can be seen for the image displacements. This seems to be a systematic effect that has also been observed in both parametric \citep{bergamini23} and free-form \citep{cha23} reconstructions, and more careful study is needed to determine the exact cause.

Visible in the top panel of Figure \ref{fig:massdist} are the predicted images that are unaffiliated with any observed images. These unaffiliated images are common features of free-form lens reconstructions despite being frequently ignored. In our model, we have a total of 58 of these, all of which can be reasonably explained. We report that 30 are distant 3rd images from sources with only 2 observed images. These images are predicted by gravitational lensing (since the number of multiple images formed for a source must always be odd) and likely unobserved in MACSJ0416 due to their weak magnification. For example, the Mothra arc only has 2 counterimages observed at the location of the arc. Our model predicts the 3rd image to form $\sim$44" away on the opposite side of the cluster with a magnification of $\sim$2 (compared to magnification at the arc location to be $\gtrsim$20). This corresponds to an apparent magnitude of $\sim$31, which is dimmer than the recent JWST observational depths of $\sim$29 \citep{diego23b}. Therefore, we predict the existence of this third image that could be potentially probed with deeper JWST exposures in future observations.

In a similar vein, 4 unaffiliated images are unobserved central images. For this case, these are likely unobserved due to the images forming too close to the cluster center, where they are superimposed by the BCGs and other nearby galaxies, making them hard to isolate. These images are also maxima in the time delay surface, and thus are also demagnified rendering them unlikely to be observed. Observations in UV filters have been suggested as a way to discover central images in lens systems \citep{perera23}, and if successful, would provide tighter constraints on inner structure of density profiles of mass structure. There are 22 unaffiliated images that are the result of critical curves ``folding'' reconstructed images into multiple copies, similar to the extraneous images found in models of SDSS J1004+4112 \citep{forestoribio22,perera24}. The cause of these critical curves are model-predicted isolated substructures at the location of the images. Since these unaffiliated images typically form very close (within $\sim 1$") to the actual observed image (and thus have very similar time delays and magnifications), they can be reasonably ignored. The remaining 2 unaffiliated images in our model form along the Spock arc, and we discuss these further in Section \ref{txt:spock}.

We note that our model fails to correctly reconstruct 3 sources, instead reconstructing too few images. Two of these, Sys14 and Sys16 as labelled in \cite{bergamini23}, are galaxy-galaxy strong lens systems (angular separation $\sim$1" for these images) about the cluster member galaxies Gal-8971 and Gal-8785, respectively \citep{vanzella17,bergamini21}. Since our model is a cluster scale reconstruction, it is not of concern that FF00 did not reconstruct these 2 sources. In order to correctly model them, we suggest a future hybrid lens model with {\tt GRALE} that explicitly includes parametric forms for the two galaxies. The other incorrect reconstruction is source 12.4 in the Warhol arc, which we discuss further in Section \ref{txt:warhol}.

\begin{figure}
\includegraphics[trim={4.2cm 0.3cm 4.5cm 0.6cm},clip,width=0.49\textwidth]{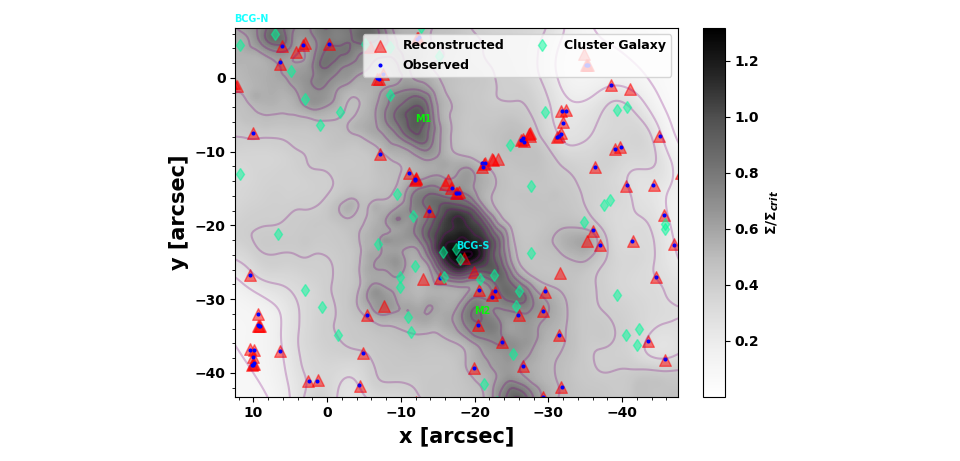}

\caption{Zoomed in view of the projected surface mass density distribution in the region surrounding BCG-S. Observed and reconstructed images are shown as blue dots and red triangles, respectively. Cluster member galaxies are identified as light green diamonds. We label two mass peaks that are not associated with a cluster galaxy as M1 and M2 in bright green. These features are potentially dark matter substructure. }
\label{fig:subst00}
\end{figure}

Figure \ref{fig:subst00} shows a zoomed in view of the mass features in the vicinity of BCG-S. We identify two mass substructures (M1 and M2) that are unaffiliated with any cluster galaxy and thus any light. Furthermore, no excess X-ray emission is detected with Chandra at their locations \citep{bonamigo17,bonamigo18}, with the X-ray profiles remaining smooth. Because of this, it is possible that these are dark matter substructures. We note that it remains to be explored if these substructures could also arise from some extremely low surface brightness cluster galaxy population. Additional deep observations would be needed to confirm the existence of such a population.

M1 is located in between the two BCGs roughly in the center of the cluster. The nearest cluster galaxy to the peak of M1 is $\sim$23 kpc away in the northeast direction. The nearest observed image is $\sim$37 kpc away. M1 contains $9.5 \pm 0.5 \times 10^{11} M_{\odot}$ within a core radius of $\sim16$ kpc. The core radius here is defined as the radius at which the density profile becomes isothermal ($d\ln\Sigma / d\ln r = -1$).  It can be argued that M1 is caused by a relatively unconstrained region in the lens plane. Substructures with these characteristics can be subject to the monopole degeneracy, where mass can be redistributed within the local region bounded by observed images without changing image positions and time delays. This degeneracy can be difficult to break without a high density of images near the substructure \citep{liesenborgs08,liesenborgs24}. Despite this, the mass substructure persists across all the free-form models generated in this study. Likewise, the location of the nearby observed image constrains the scale to which the monopole degeneracy can redistribute M1's mass. These reasons both support the existence of M1 as a dark matter substructure, but its shape and extent are less certain. 

In comparison, M2 seems to be a stronger candidate for dark matter substructure. It is smaller than M1, with $5.7 \pm 0.2 \times 10^{11} M_{\odot}$ within a radius of $\sim$8 kpc (roughly corresponding to the point at which $\Sigma$ returns to the background density). The most significant difference with M1 is its position. Despite being closer to BCG-S, the nearest cluster member galaxy (not BCG-S) is $\sim$26 kpc away while the nearest observed image lies $\sim$7.9 kpc from the mass peak. This image is a maximum in the time delay surface belonging to Sys205 ($z_s = 3.715$). Since maxima generally form close to mass peaks, this offers evidence in favor of the existence of M2. In fact, M2 mirrors a similar substructure found in Abell 1689 by \cite{ghosh23}. In both cases, the substructure forms a local peak near the BCG with observed central maxima near the vicinity of the substructure mass peak. \cite{ghosh23} argue that their substructure needs to be present in order to reconstruct the observed images near it, which we accordingly adopt for M2 due to the similarities. Furthermore, the presence of the central image so near to the peak of M2 acts as a very strong constraint on the mass distribution in that region, restricting the scale on which one can redistribute the mass of M2 with the monopole degeneracy. As with M1, M2 also persists in all the free-form models we generate.  For all these reasons, M2 is a strong candidate for dark matter substructure in the lens. We caution, however, that more data and study of the lens models in MACSJ0416 is required in order to confirm with certainty that M1 and M2 are indeed real substructures. However, in what follows, we discuss implications on models of dark matter assuming that M1 and M2 are real, in Section \ref{txt:darkmatter}. Additionally, as we describe below, M1 and M2 persist in both hybrid models, with slightly more mass than in FF00.

\subsubsection{Hybrid Models: H-Ser and H-NFW}\label{txt:hybriddensdist}

As described in Section \ref{txt:input}, we generate two hybrid models, H-Ser and H-NFW, using Sersic and NFW lens models, respectively, for the BCGs and Spock galaxies. In this section, we briefly examine the results of the two hybrid models generated for MACSJ0416, shown in the two rows of Figure \ref{fig:Hextra}. In both cases, the surface mass density profile is morphologically similar to that of FF00 on large scales. The main differences are near the BCGs and Spock galaxies, as these are where the additional parametric constraints were applied. 

For H-Ser, the mass within 200 kpc of BCG-N and BCG-S is $1.431 \pm 0.002 \times 10^{14} M_{\odot}$ and $1.486 \pm 0.002 \times 10^{14} M_{\odot}$, respectively. The mass profile around the BCGs is much more peaked in comparison with FF00, which is a result of explicitly including mass in the region with Sersic profiles. M1 has a core radius of $\sim$18 kpc with a mass of $11 \pm 0.5 \times 10^{11} M_{\odot}$, while M2 has a mass of $7.3 \pm 0.2 \times 10^{11} M_{\odot}$ within $\sim$8 kpc. In this case, both M1 and M2 are more massive than their counterparts in FF00. It should be noted that M2 features more as a mass extension from the BCG-S region rather than as a distinct substructure. This could be a side effect of the greater mass concentration at the location of BCG-S in H-Ser muting mass features in its vicinity. H-Ser finds $\Delta_{\rm RMS} =$ 0.207"; still a good fit to the data, but marginally not as accurate as FF00.

For H-NFW, the total mass within 200 kpc is $1.451 \pm 0.002 \times 10^{14} M_{\odot}$ and $1.508 \pm 0.002 \times 10^{14} M_{\odot}$ for BCG-N and BCG-S, respectively. As expected with NFW profiles, all the modeled galaxies have more spread out mass profiles than in H-Ser (see Figure \ref{fig:Hextra}). The substructures are similar to those recovered in H-Ser, with M1 having a core radius of $\sim$17 kpc and a mass of $11 \pm 0.5 \times 10^{11} M_{\odot}$ and M2 having a radius of $\sim$8 kpc and a mass of $5.0 \pm 0.2 \times 10^{11} M_{\odot}$. The $\Delta_{\rm RMS} =$ 0.206". Despite both hybrid models having low $\Delta_{\rm RMS}$, they are not as accurate as FF00.

\subsubsection{Reconstruction of the Warhol Arc}\label{txt:warhol}

\begin{figure}
\includegraphics[trim={0.1cm 1.2cm 0.1cm 1.2cm},clip,width=0.49\textwidth]{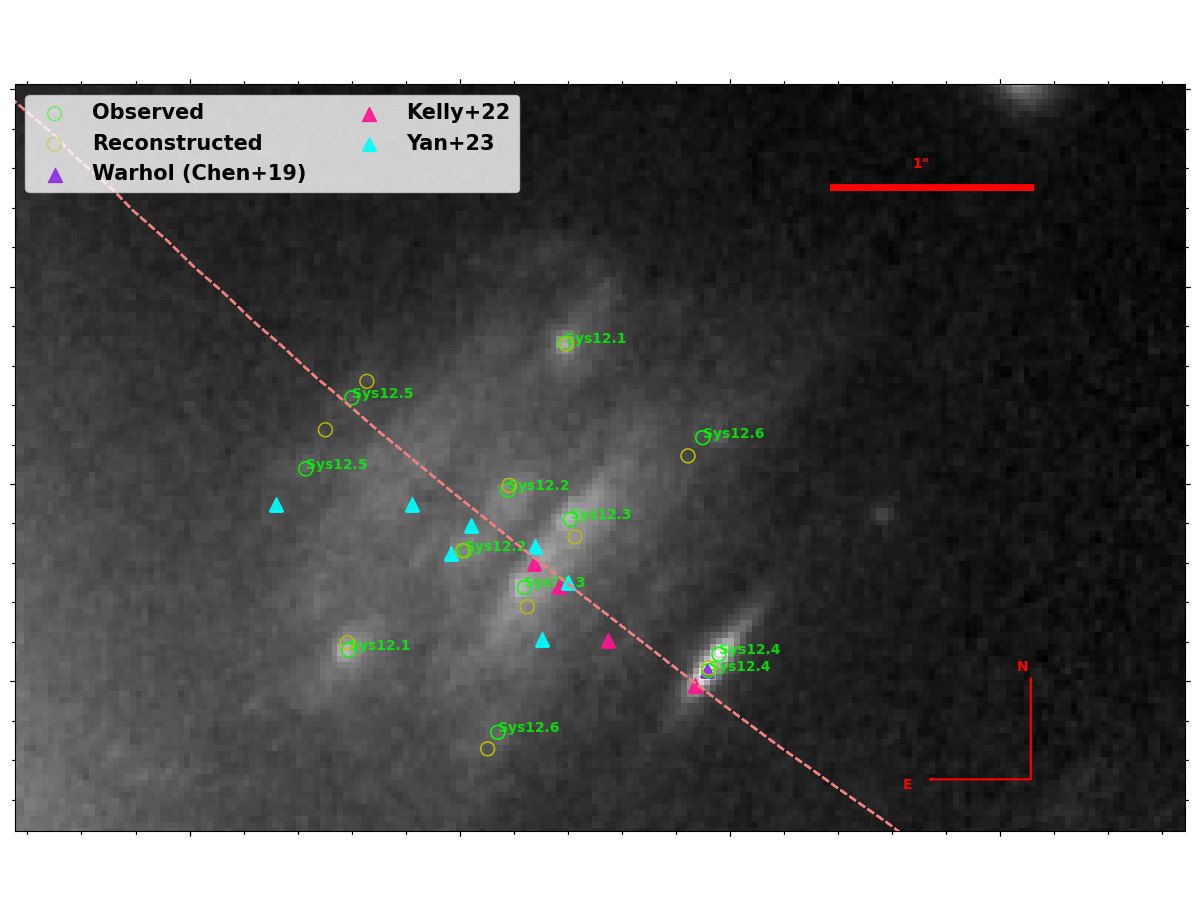}

\caption{Stacked image of the Warhol Arc combining HST F435W, F606W, and F814W, with North up and East left. The FF00 critical curve is shown as the dashed light red line. Observed and Reconstructed images for all 6 sources in the arc \citep{bergamini23} are shown bright green and gold, respectively. Sys12.4 is the pair of  observed images on the western most side of the arc (furthest right on the arc near the critical curve). The Warhol transient \citep{kaurov19,chen19} is shown in dark purple, while additional transients are shown in pink \citep{kelly22} and cyan \citep{yan23}.} 
\label{fig:warhol}
\end{figure}

An interesting arc to briefly discuss is the Warhol arc, shown in Figure \ref{fig:warhol}. It is the lowest redshift lensed source ($z_s = 0.94$) and has been host to numerous recently discovered transient events \citep{kaurov19,chen19,kelly22,yan23}. Even though the density profile in the region is not particularly interesting, the addition of 5 new multiple image sets to the dataset \citep{bergamini23} makes this system worthy of closer inspection. 

We use 6 multiply imaged sources (for a total of 12 images in the arc) to represent the Warhol arc (Sys12 in \cite{bergamini23}). To quantify the accuracy of the reconstruction of individual sources, we use the mean image separation $\langle \Delta \boldsymbol{\theta} \rangle$, defined to be the average separation in the lens plane between observed and reconstructed images. In the Warhol arc, the images are reconstructed quite well to a $\langle \Delta \boldsymbol{\theta} \rangle$ of 0.08", not including Sys12.4 as this was incorrectly reconstructed as mentioned previously. By eye, it appears that our model's critical curve finds roughly the correct midpoint (to within $\sim0.1$") of the arc that symmetrically splits most of the observed images. Its placement is also corroborated by the discoveries of numerous transients that lie along the predicted critical curve. The predominance of transients on the negative parity side of the cluster, as implied by our model, may argue in favor of wave dark matter \citep{Broadhurst2024} with de Broglie scale density perturbations arising from dark matter as an ultra-light boson. In addition, we calculate the magnification at the position of the Warhol transient \citep{kaurov19,chen19} to be $83 \pm 105$, which is consistent with previous models \citep{chen19}.

Of note is the failure to adequately reconstruct Sys12.4, the western most multiple image sources in the arc. The two observed images form very close to one another, separated by 0.1". Our predicted critical curve misses their respective midpoint (where one would expect it) by $\sim$0.1" to the South. This causes our model to only predict one image for the two observed, which forms in between them. We note that cases very similar to ours (where the predicted critical curve fails to pass through an expected symmetry point) are not uncommon \citep[e.g.][]{keeton10,diego24}, and our critical curve result lies within the image RMS for the cluster.

Whatever the reason may be for our model's inability to appropriately reconstruct Sys12.4, our model is successful at reconstructing all the other main features of the Warhol arc to excellent precision and is consistent with transient discoveries. We suggest a more detailed extended source analysis for future study to improve our model.

\subsection{Spock Arc Critical Curves}\label{txt:spock}

\begin{figure}
\includegraphics[trim={2cm 0cm 2cm 0cm},clip,width=0.49\textwidth]{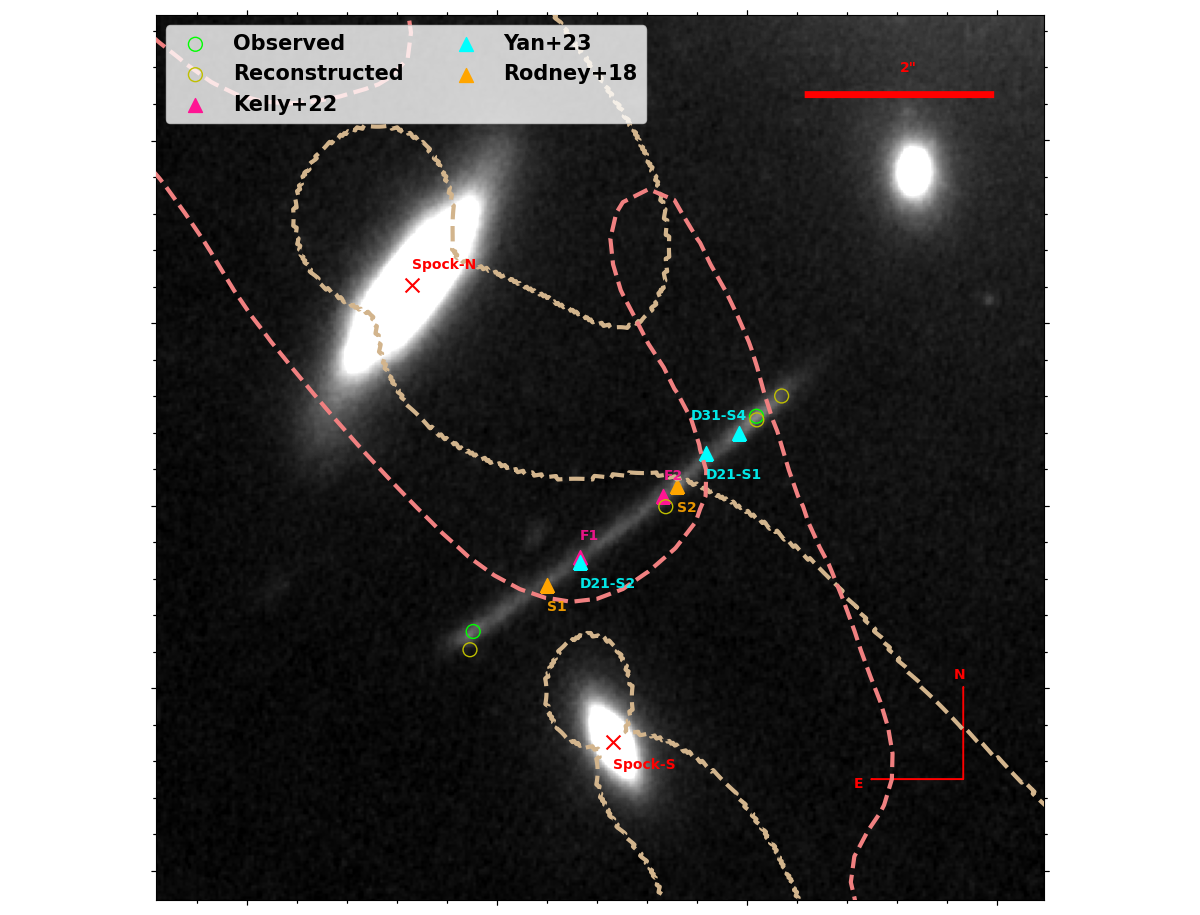}
\includegraphics[trim={5.9cm 0.35cm 5.7cm 0.35cm},clip,width=0.49\textwidth]{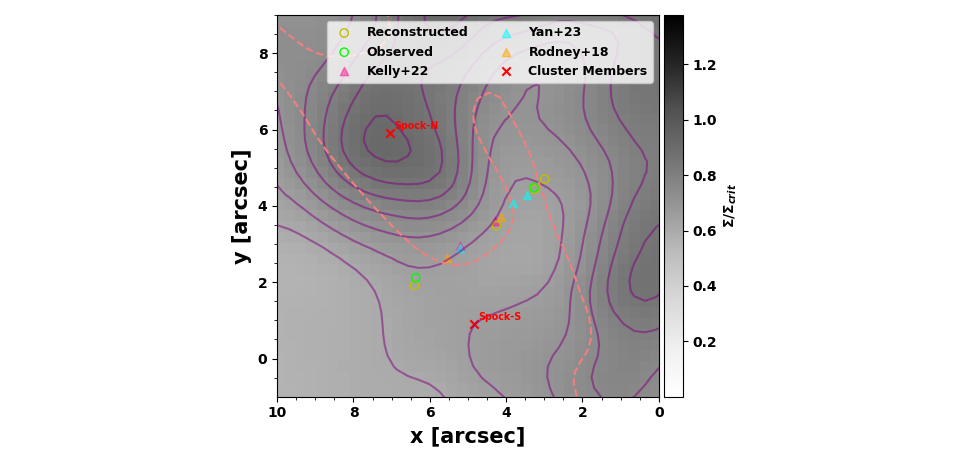}

\caption{The Spock Arc. {\it Top:} Stacked image of the arc combining HST F435W, F606W, and F814W. The FF00 critical curve is presented as the dashed light red line. Observed and Reconstructed images for the Spock arc are shown in bright green and gold, respectively. Transients are also overplotted, with S1/S2 \citep{rodney18} in orange, F1/F2 \citep{kelly22} in pink, and D21-S1/S2 and D31-S4 \citep{yan23} in cyan. The two nearby cluster member galaxies, Spock-N and Spock-S, are shown in red. The dashed light brown line is the critical curve from a recent parametric model \citep{bergamini23} shown for comparison. In the figure, North is up and East is left. {\it Bottom:} Surface mass density profile (with respect to $\Sigma_{crit}$ at $z_s =$ 1.005 for the Spock galaxy) in the region of the Spock arc. The $x$ and $y$ coordinates are with respect to the zero point of FF00. The critical curve is the same as the top panel. The field of view of this panel is the same as that of the top panel image, allowing ease of comparison of our mass distribution with the observed light.} 
\label{fig:spockCC}
\end{figure}

One of the most interesting features of MACSJ0416 is the Spock arc at $z_s = 1.005$. Figure \ref{fig:spockCC} shows the Spock Arc along with two nearby cluster member galaxies (Spock-N to the North and Spock-S to the South) and the numerous transient events recently discovered \citep{rodney18,kelly22,yan23}. These transient events are hypothesized to be bright supergiant stars that form very close to the cluster caustic, allowing them to briefly become visible during a microlensing event. This interpretation is supported by observations of the original Spock transients, S1 and S2 \citep{rodney18}, where it was concluded that the two transients likely originate from the same position in the source plane but are not ``temporally coincident'', meaning they likely did not occur at the same time. This result implies two possible explanations for S1 and S2 strictly dependent on the critical curve structure. If there is one critical curve splitting through the arc, the preferred explanation of \cite{rodney18} is a single massive Luminous Blue Variable (LBV) star undergoing two distinct surface eruptions. If there are multiple critical curve crossings (or simply a critical curve structure producing high magnification along the arc), the preferred explanation is that S1 and S2 are distinct microlensing events of two different bright stars. This latter explanation seems to be corroborated by the discovery of many more transients in the Spock Arc and simulations of the transient detection rate (Li et. al. in prep.). Despite this, most recent lens models have been unable to reconstruct a critical curve structure with sufficiently high magnification across the arc \citep{bergamini23,diego23a,diego24}, with one of the few exceptions being \cite{raney20}. 

Our FF00 model is the second lens model (after \cite{raney20}) to reconstruct a Spock arc critical curve with multiple crossings (and thus have high magnification), as shown in Figure \ref{fig:spockCC}. Specifically, our model finds 2 main crossing points on the inner part of the arc. On the western side of the arc, the critical curve passes very close to the edge of the arc, nearly forming a third crossing. The ``U'' shape of the critical curve in this region traces a mass valley west of Spock-N, which has the effect of magnifying the west side more. This result, if representative of the true nature of the critical curve of the cluster, has significant implications for the interpretation of the Spock arc. One implication is that the critical curve structure is primarily shaped by Spock-N, since no clear mass peak corresponds to Spock-S. It is also worth mentioning that these galaxies have slightly different redshifts than the BCGs, which can result in small changes to the critical curve structure \citep{rodney18,diego24}. Most importantly, this result implies that the transient events discovered in the arc are most likely microlensing events, which can place tight constraints on the abundance of supergiant stars in the Spock galaxy \citep{diego24}. Therefore, we proceed to rigorously test this result to ensure that it is robust. Appendix \ref{txt:spockunc} discusses the uncertainty in our result.

\subsubsection{Multiple Critical Curve Crossings as a Probable Explanation for Observations in the Spock Arc}\label{txt:multCC}

First, as a simple comparison, the predicted magnifications at the locations of transients are consistent with previous models. Notably, \cite{bergamini23} finds a $\mu$ of 612.6, 87.7, and 139 for the 3 observed transients in \cite{yan23}: D21-S1, D21-S2, and D31-S4, respectively. We find a $\mu$ of $647 \pm 475$, $58 \pm 44$, and $134 \pm 39$ for the same three with FF00, which is in excellent agreement. The similarity between the two models is due to their single critical curve and one of our critical curve crossings passing close to D21-S1 and equidistant from D31-S4 (see top panel of Figure \ref{fig:spockCC}) coupled with a close grazing of the arc on the Eastern side near D21-S2 by the critical curve from \cite{bergamini23}. This has the effect of producing high magnification along the arc similar to our model's two crossings.

Due to the aforementioned magnifications, we find that a critical curve crossing likely occurs near the location of D21-S1. This is further supported by the predicted large magnification of S2 ($\mu = 167 \pm 65$) which is located near D21-S1. Our model finds the Western Spock image to have $\mu = 233 \pm 37$ which is consistent with previous models \citep{zitrin13,jauzac14,caminha17,bergamini23} that also find a larger $\mu$ for the Western Spock image. If this is true, then a single critical curve model would cross the Spock arc $\sim$0.7" from the west Spock image. There are slight issues with this: (i) that the critical curve does not cross at a natural symmetry point in the arc, and (ii) that the transients on the opposite end of the arc (namely S1, F1, and D21-S2) would have lower predicted magnifications. The lower magnifications are of particular significance, as a macromodel $\mu \lesssim 40$ implies that the source star must be very bright with $M_V \sim -7$, and thus very rare, in order to be visible from a microlensing event \citep{diego24}. This can be alleviated by a special microlensing scenario whereby at least 2 lens stars contribute to the added microlensing magnification, although this is unlikely for greater distances from the critical curve \citep{diego24,palencia23}, as would be the case for only a single critical curve crossing.

\begin{figure}
\includegraphics[trim={0cm 0cm 0.6cm 0cm},clip,width=0.49\textwidth]{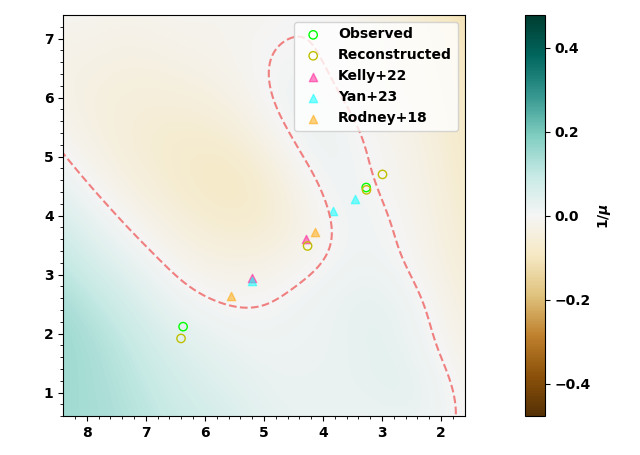}
\includegraphics[trim={0cm 0cm 0cm 0cm},clip,width=0.49\textwidth]{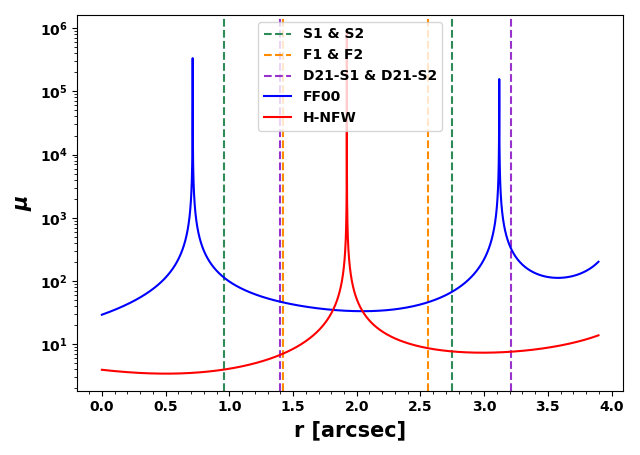}

\caption{ {\it Top:} Spock Arc magnification map. The critical curve is shown as a dashed light red line, and transients (triangles) are shown in pink (F1 and F2), cyan (D21-S1 and D21-S2), and orange (S1 and S2). The colorbar indicates the inverse of the magnification, where darker brown colors correspond to less magnified negative parity regions and darker green colors correspond to less magnified positive parity regions. Whiter regions indicate areas of high magnification. {\it Bottom:} The magnification along the Spock arc as a function of the distance $r$ from the eastern image of the Spock arc (i.e. from left to right when viewing the top panel). The model FF00 is shown in blue while H-NFW is shown in red. The spikes correspond to critical curve crossings. The vertical dashed lines represent the locations of the transients along the arc. Noteworthy is the fact that the FF00 model contains 2 critical curve crossings in the vicinity of all the transients, along with a high magnification along the whole arc. The H-NFW model, shown for comparison, features only one critical curve crossing (roughly at the symmetry point of the arc) and is roughly at least 1 order of magnitude lower in magnification along the arc. This latter scenario is unlikely to explain the high number of transients in the arc as it has a lower probability of microlensing \citep{diego24,palencia23} and low magnification implying very rare high mass lensed stars as sources. } 
\label{fig:spockmagn}
\end{figure}

The much simpler and more likely scenario is therefore that the macromodel magnification at the locations of all transients remain $\gtrsim$40, implying that the lensed stars are supergiants with $M_V \lesssim -5$, consistent with a more common blue supergiant population. This implies that the distance of the transients to the critical curve should also be smaller as this would increase the microlensing probability. Our FF00 model finds that all mentioned transients in the Spock arc have $\mu > 50$ and are within 0.7" from a critical curve crossing, which is possible due to the 2 critical curve crossings providing high magnification along the arc. To our knowledge, this is one of the only models that accounts for the high number of transients in the Spock arc. This effect is shown in Figure \ref{fig:spockmagn}. Models with a single critical curve crossing through the center of the arc are more likely to place some transients $>$1" from a critical curve (low probability of microlensing) and have low magnification along the entire arc, which is unlikely to explain the frequency and locations of the observed transients in the arc. We note that this is not universally true for models with single critical curve crossings, as these models can have close critical curve approaches of the arc contributing to high magnification \citep{bergamini23}. We conclude that our FF00 model with two critical curve crossings is an adequate representation of the Spock arc and is most consistent with observations of transients within the arc. We refer the reader to Appendix \ref{txt:spockunc} for a brief discussion on the uncertainty in this result.

As shown in the bottom panel of Figure \ref{fig:spockCC}, our reconstruction of the Spock arc produces two extraneous images: represented by two green circles, in addition to the two green circles that coincide well with two yellow circles of observed images. The most interesting aspect of these extraneous images is that both form on the negative parity side of the cluster, meaning the observed images have positive parity (see top panel of Figure \ref{fig:spockmagn} where both observed images lie in the red region). Since gravitational lensing requires counterimages to have opposite parity, these extraneous images are instead not treated as byproducts of the lens model, but rather as predicted locations of images. If we compare these locations with the Frontier Fields image of the Spock arc (see top panel of Figure \ref{fig:spockCC}), we see that the western edge of the arc does appear to have an extended light feature, which could hint that the predicted counterimage resides there. Likewise, the predicted image near the center of the arc has a small associated light feature\footnote{These two bright light features of the Spock arc are most easily seen in the bottom right panel of Figure 10 in \cite{bergamini23}.}. It should be noted that additional counterimages within the Spock arc have yet to be reported. We suggest future reexamination of the Spock arc to see if any counterimages exist at the predicted locations as predicted by FF00. If these are found, they would provide strong support for FF00.

\subsubsection{Comparison with Hybrid Models}\label{txt:spockcompa}

With our result established as a favorable model to explain the Spock arc, we now seek to test this critical curve structure by including local cluster member galaxies as explicit parametric mass components. As described in Section \ref{txt:input}, we generate 2 hybrid models: H-NFW and H-Ser using NFW and Sersic profiles, respectively, to represent the BCGs and, relevant here, the Spock galaxies, Spock-N and Spock-S. These galaxies are visible in Figure \ref{fig:spockCC}, with Spock-N larger than Spock-S \citep{tortorelli23}. The Spock galaxies are the only non-BCG cluster member galaxies that we include in these models. As mentioned in Section \ref{txt:input}, this choice is not expected to have significant bias since the cluster member galaxies do not contribute much mass to the cluster far from BCG-N \citep{bonamigo18}. 

The mass density distribution and magnification map around the Spock arc for H-NFW and H-Ser are shown in Figure \ref{fig:Hextra}, respectively. In both cases, clear mass peaks are visible representing Spock-N and Spock-S. Likewise, the total mass within 10 kpc of both galaxies in both models is consistent with what is expected from the virial mass (see Table \ref{tab:hybridN}). 

Despite this, only one critical curve crossing is found for the Spock arc in both models, passing roughly through the midpoint of the arc. $\langle \Delta \boldsymbol{\theta} \rangle$ is 0.16" and 0.24" for the observed Spock images in H-NFW and H-Ser, respectively. Both models do not reconstruct the observed images as well as FF00, which has $\langle \Delta \boldsymbol{\theta} \rangle = $ 0.11". Both models also find that the critical curve passes just beyond outer edges of the arc in addition to the main crossing at the center. This has the effect of producing predicted counterimages beyond the main observed images. While this was also seen with one of the images in FF00, in this case the images form $\sim$1.6" from the observed Spock images. This is not along the arc unlike the similar case in FF00. Most importantly, the magnification along the arc is much lower than in the case of FF00, as shown in the bottom panel of Figure \ref{fig:spockmagn}. In fact, nearly all the Spock transients in both hybrid models have magnifications $< 40$, which, as described above, is not conducive to the likely explanation of normal supergiant stars microlensed to visibility. The only exception is F2 in H-Ser due to forming close to the single critical curve crossing. 

For the main reasons of higher image RMS and low probability of microlensing, we suspect that H-NFW and H-Ser are insufficient explanations for the observations within the Spock arc. We note that the issue of low magnification can also be compensated for by a larger number density of source stars. In such a scenario, H-NFW and H-Ser could potentially better match the observations in the Spock arc, since high magnification models such as FF00 would over-predict the transient detection rate. Further monitoring of the Spock arc is therefore needed in order to more tightly constrain the transient detection rate and the Spock galaxy's initial mass function. At the current constraints from transients, however, H-NFW and H-Ser do not seem to adequately explain the observed transients as well as FF00. The reconstructions from \cite{bergamini23} and \cite{diego23a}, which both recover a single critical curve crossing, exhibit $\langle \Delta \boldsymbol{\theta} \rangle$ of 0.49" and 0.62" for the Spock arc, respectively, compared to our FF00 model's $\langle \Delta \boldsymbol{\theta} \rangle = $ 0.11". This further strengthens our conclusion that a multiple critical curve crossing structure is needed to explain the Spock arc. 

\subsubsection{$M/L$ Ratios of the Spock Galaxies}\label{txt:spockML}

Since {\tt GRALE} does not include any cluster galaxy information as input, it presents a light-agnostic view of the mass distribution in the cluster. Therefore, unlike in parametric or hybrid models, it is necessary to check if enough mass is produced in the model to account for the stellar contribution in cluster member galaxies. This is a strong requirement. In the case of Spock-S, this is of particular interest since our model does not find an obvious mass concentration there. If there is insufficient mass to account for the light of Spock-S, then this can indicate that the 2 critical curve crossings of the Spock arc are influenced strongly by the model's failure to reconstruct Spock-S.

To check this, we calculate the mass-to-light ($M/L$) ratios for Spock-N and Spock-S. Luminosities are calculated in HST F160W since the elliptical Spock galaxies are brighter in infrared (IR) filters. \cite{tortorelli23} measures the total F160W AB magnitude $m_{160W}$, allowing us to easily measure the luminosity:
\begin{equation}
   \log_{10} \left( \frac{L}{L_{\odot}} \right) = 0.4\left( M_{sun} - m_{160W} - 5 + 5\log_{10}\left(D_d (1+z_d)^2\right) \right),
\end{equation}
where $M_{sun}$ is the absolute magnitude of the Sun, corresponding to 4.60 in F160W \citep{willmer18}. The background subtracted mass can be easily calculated from our mass model, which we find to be $6.5 \pm 1.9 \times 10^{10} M_{\odot}$ and  $6.0 \pm 0.4 \times 10^{11} M_{\odot}$ within 10 $R_e$ for Spock-S and Spock-N, respectively. This corresponds to $M/L$ of $3.59 \pm 1.06$ and $6.48 \pm 0.43$ in solar units for Spock-S and Spock-N, respectively. This is consistent with results from \cite{humphrey06}, who find a $M/L$ range of $\sim$3-8 for a sample of 7 elliptical galaxies at this radius. It is also in agreement with $M/L$ relation with mass measured from the SAURON project \citep{cappellari06}. This result shows our model is successfully able to reconstruct sufficient mass at the locations of the Spock galaxies, even though they are not included as prior constraints in the model. Since Spock-S has sufficient mass and does not significantly contribute to the critical curve crossings, we conclude that the 2 critical curve crossings can be explained primarily by the structure around Spock-N, and are likely not a result of the model failing to reconstruct the mass around Spock-S.

\subsubsection{Possible Transient Counterimages in the Spock Arc}\label{txt:spockID}

The last consideration is the controversy surrounding whether or not any of the transients are counterimages of one another, as detailed in Section \ref{txt:input}. The 3 transient counterimage hypotheses we test are S1/F1 and S2/F2 (FF11, shown in the top row of Figure \ref{fig:FFextra}), S1/S2 and F1/F2 (FF12, shown in the second row of Figure \ref{fig:FFextra}), and D21-S1/D21-S2 which are included in addition to the scenarios of FF11 (FF11+D, shown in the third row of Figure \ref{fig:FFextra}) and FF12 (FF12+D, shown in the bottom row of Figure \ref{fig:FFextra}). All 4 models have slightly larger $\Delta_{\rm RMS}$ and mean image separations $\langle \Delta \boldsymbol{\theta} \rangle$ in the Spock arc than FF00.

In the cases where S1/F1 and S2/F2 are tested as counterimages (FF11 and FF11+D), we find that both cases find only a single critical curve crossing at roughly the midpoint of the arc (see Figure \ref{fig:FFextra}). If S1/F1 and S2/F2 were counterimages, then both pairs require a critical curve to pass between them, since counterimages in gravitational lensing must be split by a critical curve. Thus, the reconstructed single critical curve does not properly reproduce a lensing structure that would imply that S1/F1 and S2/F2 are counterimages. This is somewhat surprising due to FF00 (which did not include transient counterimages) producing two critical curve crossings near these transients. We therefore conclude that the Spock S transients \citep{rodney18} are most likely not counterimages of the Spock F transients \citep{kelly22}. 

The cases with S1/S2 and F1/F2 as counterimages (FF12 and FF12+D) are more interesting as their counterimage requirement is a single critical curve passing through the arc midpoint, which is what we find in both models. $\langle \Delta \boldsymbol{\theta} \rangle$ is $\sim$0.28" and $\sim$0.12" for F1/F2 and S1/S2, respectively. In FF12+D, $\langle \Delta \boldsymbol{\theta} \rangle$ for D21-S1/D21-S2 is $\sim$0.5". This is very similar to what was found in FF11+D. While the reconstructions in this case are acceptable, they significantly fail at reconstructing appropriate magnification for these sources. From the argument in Section \ref{txt:multCC}, the macromodel magnification at the location of all the transients needs to be $\gtrsim$40 to be consistent with a common blue supergiant. However, in FF12 and FF12+D $\mu \lesssim 20$ at the location of each transient. This is categorically lower than those recovered in the hybrid models. For the same reasons then, we also conclude that these models provide insufficient explanations for the transients in the arc. 

The most likely model that we generate in this work is FF00 with 2 critical curve crossings allowing for a high probability that all the observed transients are affected by microlensing.

\subsubsection{Upper Limits on the Luminosity of Lensed Stars}\label{txt:spockstars}

With FF00 established as a strong candidate to effectively describe the Spock arc, we now seek to derive constraints on the nature of the transients in the arc. As mentioned in previous works \citep{rodney18,kelly22,yan23,diego24}, the transients in the Spock arc are most likely highly magnified supergiant stars. To estimate the luminosity, all we must do is correct for magnification at the position of each transient to derive the intrinsic magnitude of the transient without the lens. We apply this to the recently discovered transients D21-S1, D21-S2, and D31-S4 \citep{yan23} since they have photometric measurements from JWST F200W, which we assume to be at their peak magnitude. We note that while a more robust calculation of the luminosity of lensed stars can be done with SED fitting \citep[e.g.][]{chen19,diego23b}, our results serve as a good first order check that the lens model does not suggest unusual results. The macromodel magnifications for D21-S1, D21-S2, and D31-S4 are previously mentioned in Section \ref{txt:multCC}.

With photometry of the 3 transients from \cite{yan23}, we are able to calculate the observed flux $f_{\nu,\rm obs}$. We can then calculate the intrinsic magnitude $m_{\rm exp}$ (in AB) of each transient by correcting for the predicted magnification $\mu$ by FF00 at each one's location:
\begin{equation}\label{eq:mexp}
    m_{\rm exp} = -2.5\log_{10}\left( \frac{f_{\nu,\rm obs}}{\mu} \right) - 48.60.
\end{equation}
We can then use the distance modulus to estimate the luminosity of the transient, similar to what we did in Section \ref{txt:spockML}. This calculation yields a strict upper limit of the luminosity since we do not include any likely microlensing effect and we assume $f_{\nu,\rm obs}$ is at peak brightness. We find upper limits in JWST F200W of $2.1 \pm 0.3 \times 10^5 L_{\odot}$, $1.7 \pm 0.3 \times 10^6 L_{\odot}$, and $9.0 \pm 1.4 \times 10^5 L_{\odot}$ for D21-S1, D21-S2, and D31-S4, respectively. These are broadly consistent with B8V or O3V stars \citep{meena22}. In reality, due to the high probability of microlensing, these transients likely have a much lower luminosity. For these observations, a depth of $\sim$29 was achieved \citep{yan23}, corresponding to a required magnification of $\sim10^4$ for this type of star to be visible \citep{meena22}. This is roughly an increase of 100 times on top of the macrolens magnification, which is easily achievable with microlensing. Repeating the calculation with this magnification to roughly account for microlensing gives luminosities consistent with O and B type stars. From this quick calculation, we find that FF00 suggests that the transient stars in the Spock arc are most likely massive blue main sequence stars or blue supergiants, which is in agreement with previous analyses of transients in the Spock arc \citep{rodney18,diego24}. Since these transients are only observed in one epoch, we are unable to constrain their variability. Future studies of the numerous transients in this arc will be able to use our model to more tightly constrain properties of the lensed stars, source galaxy, and the microlens density.

\subsection{Millilensing of Mothra}\label{txt:mothra}

\begin{figure}
\includegraphics[trim={5.1cm 0.1cm 5.1cm 0.1cm},clip,width=0.49\textwidth]{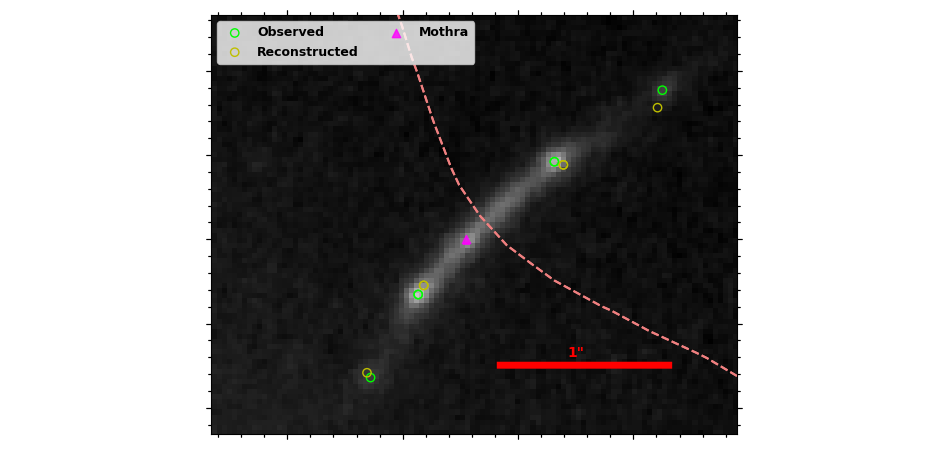}

\caption{Stacked image of the Mothra arc combining HST F435W, F606W, and F814W. Observed and reconstructed images are shown in bright green and gold, respectively. The FF00 critical curve is shown as a light red dashed line. The Mothra transient \citep{diego23b} is shown as a magenta triangle.} 
\label{fig:mothra}
\end{figure}

Here, we focus on the Mothra arc at $z_s = 2.091$ (see Figure \ref{fig:mothra}), whose name comes from the specific transient ``Mothra'' \citep{diego23b}. Mothra is well studied, with SED fitting finding that it likely consists of a binary pair of two supergiants \citep{diego23b}. Interestingly, Mothra has been visible in HST since 2014, with no confident counterimage discovered as of this work. These have led to different interpretations of the nature of Mothra, with the most likely explanation being a case of millilensing by a $\geq10^4 M_{\odot}$ substructure near the position of Mothra \citep{diego23b}. Microlensing is disfavored because of the $>8$ year flux anomaly with no counterimage \citep{diego23b}. Millilensing by substructures close to the macrolens critical curve is an emerging frontier \citep{venumadhav17,dai18,dai20a,williams24}, with the case of Mothra offering a unique opportunity to test cosmological models. Currently, the nature of Mothra's millilens remains relatively unconstrained, with upper limits on millilens mass ranging from $\sim10^6 M_{\odot}$ \citep{diego23b} to $\sim10^9 M_{\odot}$ \citep{abe23}. The relative position and size of the millilens are also unknown. To study this case of millilensing, we develop a statistical framework to place upper limits on the mass, size, and location of the millilens. 

\subsubsection{Statistical Inference of Millilens Parameters}\label{txt:mothrainference}

We adopt a Metropolis-Hastings algorithm to optimize the parameters of the millilens. This first requires several considerations in order to build a realistic posterior to sample in the algorithm. In general, we build the likelihood using the assumptions that the millilens provides sufficient magnification at Mothra and that it does not significantly perturb the Mothra arc.

First, since \cite{diego23b} uses SED fitting to identify Mothra as a binary pair of supergiants, we estimate the required magnification at the position of Mothra. This magnification needs to be achieved by the millilens perturbation to the macrolens.  The best fitting SED from \cite{diego23b} consists of a hot ($T_{\rm \rm eff} \approx 14000$ K) and cool ($T_{\rm eff} \approx 5000$ K) star. In $V$ band, the observed apparent magnitude of Mothra is $\sim$28.85, which we can convert in AB magnitudes to a corresponding $f_{\rm \nu,\rm obs}$. In this filter, the hotter star dominates the SED. This hotter star is consistent with a blue supergiant. Because of this, we adopt a prototypical absolute magnitude for Mothra equivalent to that of fiducial blue supergiant Rigel: $M_V = -7.84$ \citep{przybilla06}. This corresponds to an intrinsic magnitude $m_{\rm exp}$ of $\sim$38.29, which is the magnitude of Mothra if there were no lensing. With these two measurements, we can rearrange Equation \ref{eq:mexp} to solve for the required magnification for Mothra $\mu_m$ which we find to be 5995. This is broadly consistent with expectations described by \cite{diego23b}. 

The FF00 absolute magnification at the location of Mothra is $238_{\rm -64}^{\rm +465}$ (uncertainties are the 68\% confidence intervals), which is on the same order as the magnification of 885 found by \citep{diego23b}, and larger than the magnification of 32.5 found by \citep{bergamini23}. These are roughly an order of magnitude below what is required assuming a blue supergiant source, meaning a millilens is needed in order to increase the magnification to $\mu_m\sim 6000$. The millilens is treated as a perturbation on top of the macrolens model, meaning that it should not change the overall shape or magnification of the Mothra arc. To minimize the change in the shape of the arc, we include two terms in the likelihood function. The first simply requires that the positions of the reconstructed Mothra images $\boldsymbol{r}_{m,\rm rec}$ are minimally displaced from the observed Mothra images $\boldsymbol{r}_{m,\rm obs}$. This is the same requirement as used in the source position optimization (see Equation \ref{eq:galposterior}). We note that we only consider the two inner Mothra images (source 202.2 in \cite{bergamini23}) since the outer Mothra images are far enough away that they are largely unaffected by the presence of the millilens. The second term minimizes the millilens distortion of the cluster critical curve. We quantify this as $\xi = \max(\boldsymbol{r}_{m,\rm CC} - \boldsymbol{r}_{\rm mac,CC})$, where $\boldsymbol{r}_{m,\rm CC}$ and $\boldsymbol{r}_{\rm mac,CC}$ refer to the millilens and macrolens cluster critical curve positions, respectively. The presence of the millilens distorts the cluster critical curve, causing the parity and the magnification of the arc to lose its symmetry. Preservation of this is motivated by the fact that the Mothra arc is roughly the same brightness on either side of the cluster critical curve.

Finally, we minimize the millilens effect on the magnification of the Mothra images for the same reasons as for the critical curve. We apply this for both the inner Mothra images which have macrolens magnifications of $\mu_{\rm mac,+}$ and $\mu_{\rm mac,-}$ for the positive and negative parity images, respectively. Accordingly, the inclusion of the millilens has magnifications of $\mu_{m,+}$ and $\mu_{m,-}$.

Combining these, and assuming each millilens model property has a roughly Gaussian distribution, we use the following likelihood function in our Metropolis-Hastings optimization of the millilens:
\begin{multline}\label{eq:mothrapost}
    \ln \mathcal{L}\left(\boldsymbol{\theta}\right) = -\frac{1}{2} \sum \left[
    \left(\frac{\mu_{\theta} - \mu_m}{\sigma_{\mu}}\right)^2 + 
    \left(\frac{\boldsymbol{r}_{m,\rm rec} - \boldsymbol{r}_{m,\rm obs}}{\sigma_r}\right)^2 \right. \\ 
    + \left. \left(\frac{\mu_{m,+} - \mu_{\rm mac,+}}{\sigma_{\pm}}\right)^2 + \left(\frac{\mu_{m,-} - \mu_{\rm mac,-}}{\sigma_{\pm}}\right)^2 + \left(\frac{\xi}{\sigma_{\xi}}\right)^2 \right].
\end{multline}
Here, $\mu_{\theta}$ is the model magnification including the millilens and its respective parameters $\boldsymbol{\theta}$. We use $\sigma_{\mu} = 600$ for the model uncertainty on $\mu_{\theta}$, assuming the identification of Mothra as a blue supergiant binary is robust. This corresponds to an uncertainty on the absolute magnitude of $\sim0.1$, which is reasonable for a prototypical blue supergiant. We set $\sigma_r = 0.04"$ to be equivalent to the astrometric precision of HST, and $\sigma_{\pm} = 1$ for the magnification uncertainty at the Mothra images' location. This is purposefully smaller than $\sigma_{\mu}$ since by construction these should be minimally distorted by the millilens. The uncertainty of the cluster critical curve distortion is $\sigma_{\xi} = 0.1"$, conservatively chosen as twice the image RMS for the Mothra images. For this analysis, we adopt an uninformed prior.

With the likelihood function established, we seek to minimize equation \ref{eq:mothrapost} by sampling $\boldsymbol{\theta}$ for various millilens models. Since each $\boldsymbol{\theta}$ is distinct for different millilens models, we measure the mass and core size for each best fitting millilens model from the optimization in order to compare across different models. 

\subsubsection{Millilens Models}\label{txt:millilensmodels}

\begin{figure*} 
\begin{multicols}{2}
    \includegraphics[trim={4.1cm 0.35cm 4.1cm 0.35cm},clip,width=\linewidth]{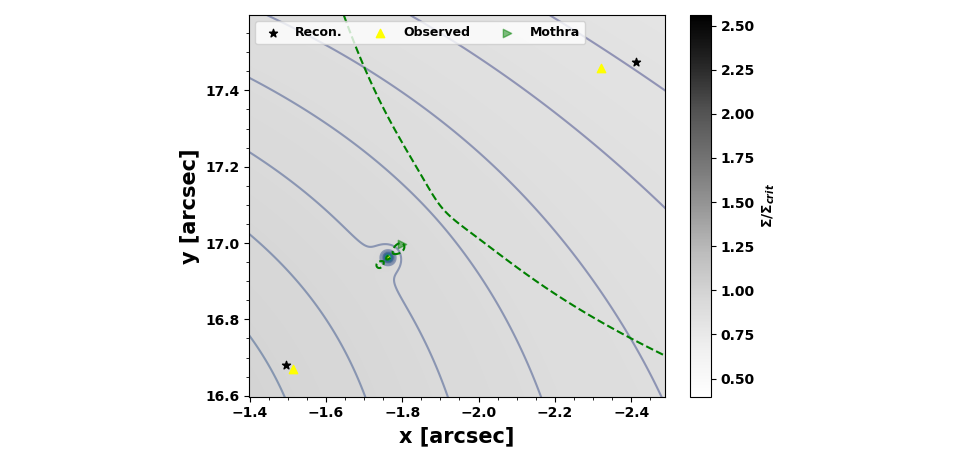}\par 
    \includegraphics[trim={0cm 0cm 0cm 0.0cm},clip,width=\linewidth]{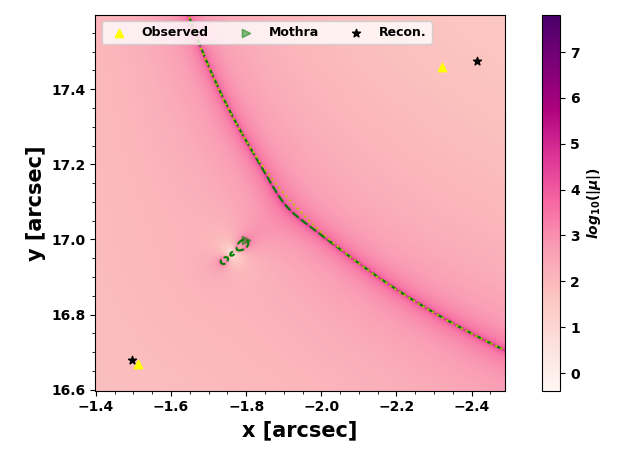}\par  
    \end{multicols}
\begin{multicols}{2}
    \includegraphics[trim={0cm 0cm 0cm 0cm},clip,width=\linewidth]{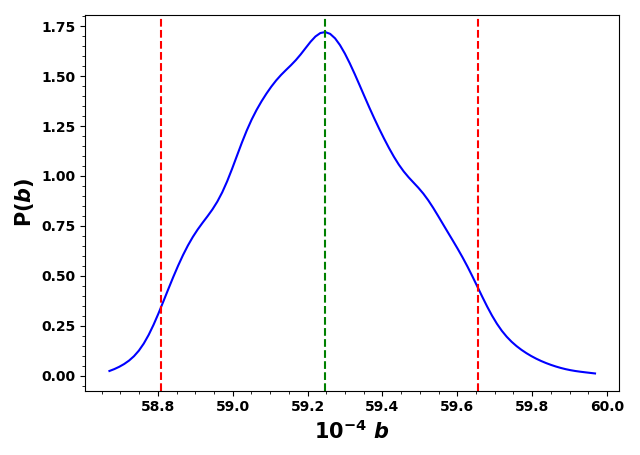}\par 
    \includegraphics[trim={0cm 0cm 0cm 0cm},clip,width=\linewidth]{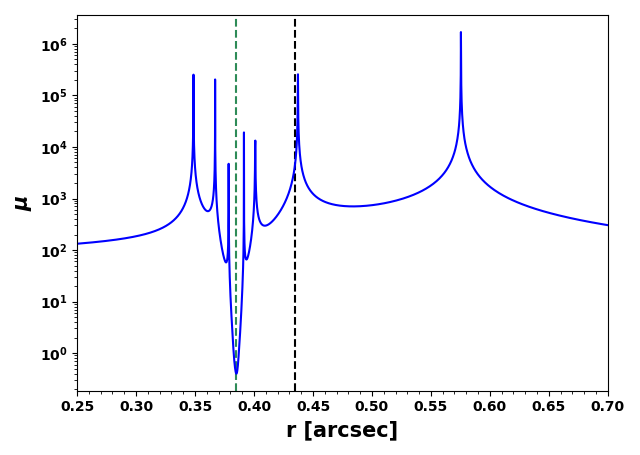}\par  
\end{multicols}
\caption{ Summary plots for the best fitting  T-strip millilens model in the Mothra arc 0.05" from the position of Mothra. {\it Top Left:} Projected surface mass density distribution in the Mothra arc relative to $\Sigma_{crit}$ at $z_s = 2.091$. Yellow triangles and black stars refer to the observed and reconstructed inner Mothra arc images, respectively. The green triangle indicates Mothra. The critical curve for is shown as a dashed green line, with the millilens critical curve clearly visible in the vicinity of Mothra. {\it Top Right:} Same as the top left plot, but with the magnification plotted instead of surface mass density. Additionally, the yellow dotted line indicates the macrolens cluster critical curve without the inclusion of the millilens. We note the small distortion of the millilens model cluster critical curve with the macrolens caused by the addition of the millilens model. {\it Bottom Left:} The posterior distribution for the T-strip parameter $b$. The green dashed line indicates the mode of the posterior. The red dashed lines indicate the 95\% credible interval. This model finds $b = 59.25^{+0.44}_{-0.41} \times 10^{-4}$. {\it Bottom Right:} The magnification $\mu$ along the Mothra arc as a function of the distance $r$ from the easternmost inner Mothra arc image (leftmost observed image in Top plots). The green and black dashed lines indicate the position of the millilens and Mothra. The magnification spikes indicate crossings of the critical curve, with the cluster critical curve represented by the spike at $r \sim 0.6$". The millilens magnification contribution is clearly seen at $r \sim 0.4$".}
\label{fig:millilens}
\end{figure*}

For this work, we only consider two models for the millilens. In general, the dense cluster environment in which the Mothra arc resides implies that millilens subhaloes will be tidally truncated with steep outer density profiles \citep{williams24}. Therefore, we restrict chosen models to those allowing for such behavior. To save computational resources, we only use lens models with an analytical form. Furthermore, the usage of two millilens models offers a view of the systematic uncertainty from the millilens.

The first is a circular power-law potential derived from {\tt alphapot} \citep{kee11}:
\begin{equation}\label{eq:alphapot}
    \psi(x,y) = N\left(s^2 + x^2 + y^2\right)^{\frac{\alpha}{2}},
\end{equation}
where $N$ is the normalization, $\alpha$ is the power-law exponent\footnote{Not to be confused with the deflection angle.}, and $s$ is the core radius. The advantage of this model is that the core radius (one of the properties of the millilens we wish to constrain) is built into the model. Similarly, the presence of $\alpha$ allows for flexibility in the millilens density slope.

The second model we consider is the tidally stripped lens potential used by \cite{williams24}, dubbed ``T-strip'':
\begin{equation}\label{eq:tstrip}
    \psi(r) = N\frac{b}{a-b}\left( a\ln\left(a+r\right) - b\ln\left(b+r\right) \right),
\end{equation}
where $N$ is again the normalization, and $a$ and $b$ are dimensionless constants. The surface mass density profile $\kappa = \frac{1}{2}\boldsymbol{\nabla}^2\psi$ falls steeply as $r^{-3}$ for the outer profile, thus mimicking a tidally stripped subhalo. Since this model is designed to explicitly represent a tidally stripped subhalo, we define T-strip to be our fiducial model.

In general, we define the core radius of these two models to be the radius at which $d\ln\Sigma / d\ln r = -1$, as this is where the density profile is approximately isothermal. The total mass of each millilens is calculated within this core radius. We note that our definition of core radius is different than the core radius $s$ used in {\tt alphapot}. 

One minor complication with this analysis is the presence of lensing degeneracies. In general, any lens model suffers from degenerate lens solutions that can equally fit the observational data. For this specific case of millilensing in Mothra, lensing degeneracies primarily appear in the determination of the best fitting model parameters $\boldsymbol{\theta}$ as there is no unique set of parameters which minimizes $\ln \mathcal{L}\left(\boldsymbol{\theta}\right)$. While this ideally can be alleviated by an increase in observational constraints, for the purposes of our analysis, we choose to break the lens degeneracy by fixing specific parameters of our lens models. For {\tt alphapot}, we fix $N = 10^{-14}$ and $s = 0.004$", leaving $\alpha$ as the sole free parameter for this model. For T-strip, we fix $N = 3.5 \times 10^{-13}$ and $a = 0.005$ (smaller than \cite{williams24} to represent a very compact subhalo), leaving $b$ as the sole free parameter for this model. In both cases, the chosen value for $N$ is rather arbitrarily chosen to ensure that the millilens density profile converges to the local macrolens density profile far away from the millilens. Fixing the core radius $s$ in {\tt alphapot} no longer allows us to constrain the core size with this model, but instead allows us to constrain the density slope. The chosen core radius of 0.004" is equivalent to $\sim$22 pc \footnote{Using our definition of core radius at the isothermal radius, this corresponds to $\sim$12 pc. This remains constant for different values of $\alpha$.}, which is on the smaller end of expected sizes of subhalos \citep{faisst22,williams24}. Fixing $a$ in T-strip does not come with this restriction, so we are able to derive constraints on mass and core radius with this model. Furthermore, since this model is more representative of tidally stripped subhalos, we quote the constraints from this model as the main results of this analysis, relegating {\tt alphapot} as a comparison model for consistency. 

The choice to fix parameters has two primary motivations. The first is that it saves significant computation time. Since the millilens is quite small (on the order of milliarcseconds), the required resolution of the lens model needs to be increased to 0.0014 arcsec per pixel (from 0.282 arcsec per pixel). This consequently increases the computation time per sample in the MCMC. With only one parameter of interest, the computation time decreases. The second is to manually break the lens degeneracy. While at first glance it may seem that we are sacrificing complexity, this is not the case. Since the parameters for the lens models are degenerate, they yield similar mass models. This means that for a different set of parameters, roughly the same mass and core size millilens is produced. Because of this, fixing a parameter is preferable to ensure that the MCMC is able to converge the likelihood to a minimum. We also note that fixing parameters significantly reduces the statistical uncertainty, hence rendering systematic effects as the best proxy for the uncertainty.

In summary, we have two millilens models: T-strip and {\tt alphapot}. We use a Metropolis-Hastings algorithm to sample the posteriors for $b$ and $\alpha$ for T-strip and {\tt alphapot}, respectively, using the likelihood function in equation \ref{eq:mothrapost}. Lastly, since in principle the millilens can be located anywhere in the lens plane, we repeat the analysis at varying positions from Mothra along the arc. We only consider positions at increasing distances from the cluster critical curve, as approaching the cluster critical curve will significantly distort the parity of the arc. Based on the rough size of each millilens, we adopt an uncertainty on each chosen millilens position of $\sim$0.01". At each position, we measure the best fitting mass for both models (and core size for T-strip) for 6000 samples in the MCMC. As an example, the best fit lens model for T-strip at a position of 0.05" from Mothra is shown in Figure \ref{fig:millilens}.

\subsubsection{Millilens Results}\label{txt:millilensresults}

\begin{figure}
\includegraphics[trim={0cm 0cm 0cm 0cm},clip,width=0.49\textwidth]{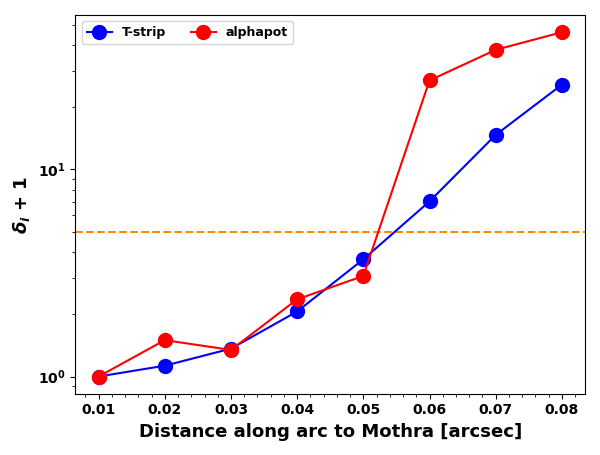}
\includegraphics[trim={0cm 0cm 0cm 0cm},clip,width=0.49\textwidth]{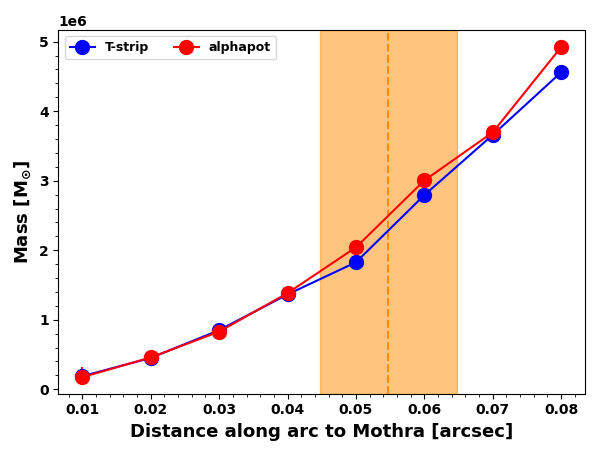}
\includegraphics[trim={0cm 0cm 0cm 0cm},clip,width=0.49\textwidth]{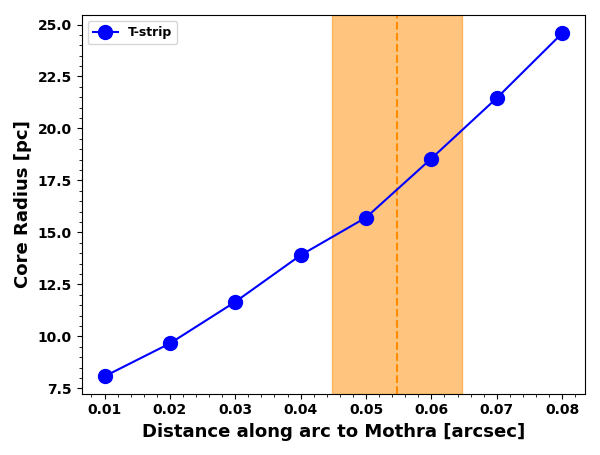}

\caption{Value of $\delta_i$ ($\Delta$AIC) for T-strip and {\tt alphapot} for models at different distances $r$ from Mothra ({\it Top}). $\delta_i$ is calculated relative to the best model for each millilens. We expect $\delta_i$ to significantly increase for $r < 0.01$", however, we do not examine this regime due to computational expense. The dashed orange line denotes $\delta_i = 4$, which we adopt as the upper limit condition on $r$. We assume that these plots depict rough interpolations of $\delta_i$ with $r$. We also show the mass ({\it Middle}) and core radius ({\it Bottom}) of the best fitting millilens as a function of $r$. The dashed orange line corresponds to the upper limit on $r$ and the intersection with both curves gives the upper limit on mass and core radius. The orange shaded region indicates the uncertainty on the upper limit on $r$ of 0.01", defined by the rough size of the millilens. We do not include {\tt alphapot} in the bottom plot since its core radius was fixed.  } 
\label{fig:mothraresults}
\end{figure}

Figure \ref{fig:mothraresults} shows main results of the analysis presented in Section \ref{txt:millilensmodels} for T-strip and {\tt alphapot} at positions along the arc ranging from 0.01" to 0.08". Comparison of the two models in the middle panel of Figure \ref{fig:mothraresults} shows the rough uncertainty on the best fitting millilens mass. To evaluate the strength of each model at different positions, we utilize the Akaike Information Criterion (AIC):
\begin{equation}\label{eq:AIC}
    AIC = 2K - 2\ln\hat{\mathcal{L}}(\boldsymbol{\theta}),
\end{equation}
where $K$ is the number of model parameters and $\hat{\mathcal{L}}(\boldsymbol{\theta})$ is the maximized likelihood function of equation \ref{eq:mothrapost}. While not the most robust statistical metric, the AIC allows us to find constraints roughly to the order of magnitude. This is motivated by the fact that for both T-strip and {\tt alphapot}, the best fitting millilens mass and AIC are roughly similar to one another (see Figure \ref{fig:mothraresults}). To compare each model at different distances, we compare each model AIC ($AIC_i$) to the minimum from the millilens models ($AIC_{\rm min}$) with $\delta_i = AIC_i - AIC_{\rm min}$. We employ the standard evaluation metric of $\delta_i > 4$, corresponding to substantially less support for the $i$th model since the relative likelihood between models drops to 0.14. With this, we can set upper limits in two stages: (1) defining the upper limit on millilens position where $\delta_i > 4$ (orange line in top panel of Figure \ref{fig:mothraresults}), (2) finding the upper limit on mass and core radius at the upper limit on millilens position (orange line in middle and bottom panels of Figure \ref{fig:mothraresults}). We emphasize that we assume AIC values between each sampled millilens position are roughly interpolated, which appears to be a realistic assumption due to the stable correlations of mass and core radius with millilens position.

Following this statistical analysis, we find the upper limit on millilens position to be 0.055" and 0.052" for T-strip and {\tt alphapot} respectively, with positional uncertainty of 0.010" as described in Section \ref{txt:millilensmodels}. Beyond this distance along the arc, the best fitting millilens mass to contribute magnification at Mothra distorts the cluster critical curve significantly, causing $\delta_i$ to blow up. In general, $\delta_i$ increases similarly for both models up to the upper limit, indicating that both models are consistent in explaining the millilens structure at different positions. Since T-strip is our defined fiducial model, we adopt 0.055" as the upper limit on millilens position. 

At this position, we calculate the upper limit on the millilens mass to be $2.29^{+0.91}_{-0.71} \times 10^6 M_{\odot}$ and $2.50^{+0.83}_{-0.81} \times 10^6 M_{\odot}$ for T-strip and {\tt alphapot} respectively. Given the uncertainty in millilens position, these upper limits are quite consistent with one another. In fact, as shown in Figure \ref{fig:mothraresults}, the mass for both T-strip and {\tt alphapot} follow closely with $r$. With T-strip, we adopt the upper limit on mass to be $2.29^{+0.91}_{-0.71} \times 10^6 M_{\odot}$. This is consistent with the upper limit of $2.5 \times 10^6 M_{\odot}$ found by \cite{diego23b}, and in tension with the upper limit of $1.4 \times 10^9 M_{\odot}$ from \cite{abe23}.

Correspondingly, the upper limit on core radius from T-strip is $17.1^{+2.8}_{-2.3}$ pc. As mentioned previously, we fixed the core radius parameter in {\tt alphapot} as $s \sim 22$ pc, and therefore do not derive constraints with it. With our definition of core radius at the isothermal density slope, the {\tt alphapot} core radius is $\sim$12 pc, in agreement with the upper limit from T-strip. For {\tt alphapot}, we constrain $\alpha \sim 0.0778$ at the upper limit, consistent with a steep outer density profile. A core radius upper limit has yet to be established for the Mothra millilens, and we study its implications in Section \ref{txt:darkmatter}.

As can be seen in Figure \ref{fig:mothraresults}, $\delta_i$ converges for our two models at 0.01" from Mothra. We expect $\delta_i$ to increase substantially at even closer distance to Mothra $r < 0.01$". This is because for masses smaller than $\sim 10^4 M_{\odot}$ (the rough millilens mass in this regime), millilens critical curves may not form, so the magnification at Mothra would be insufficient. Furthermore, the millilens needs to be $\gtrsim10^4 M_{\odot}$ in order to have sustained magnification over 7 years (as observed) with reasonable velocities and orientations of its trajectory relative to the caustic \citep{diego23b}. This means that the lower limit on the millilens position exists somewhere within 0.01" from Mothra. We are unable to probe this regime due to computational expense as we would need to drastically increase the resolution. Nevertheless, if we assume that our best fitting millilens position (0.01") is close to the global minimum for this exercise, we estimate the best fitting mass and core radius to be $1.89^{+0.02}_{-0.01} \times 10^5 M_{\odot}$ and $8.08^{+0.01}_{-0.01}$ pc, respectively.

For this analysis, we find that for both millilens models the results are consistent with our initial assumption that the millilens subhalo be tidally truncated. As a quick test of this, we repeat the analysis for an NFW profile (equation \ref{eq:nfw}) placed at $r = 0.03$"\footnote{Here, the choice of position of the NFW is arbitrary. Since this is a test of the tidal truncation assumption, it can be repeated at any position and compared with T-strip and {\tt alphapot}.}. The mass of the NFW profile does not converge at far distances, and so is a good representation of a millilens model that is not tidally truncated. We find for this model that the AIC increases significantly to 121 ($\delta_i \sim 113$), roughly an order of magnitude increase from T-strip and {\tt alphapot}. The culprit for this is the fact that the excess mass at further distances from the millilens contributes to increased distortion of the cluster critical curve and greater change in magnification at the positions of the Mothra images. Because of this, we conclude that the millilens is most likely tidally truncated.

Based on the constraints of mass and core radius, the identity of the millilens appears to be consistent with a small globular cluster. At this size, a globular cluster is potentially bright enough to be visible with JWST, but remains undetected with JWST. This interpretation is consistent with the one reached by \cite{diego23a}, who estimate a $\sim20$\% probability that a globular cluster aligns properly to provide the necessary magnification for Mothra. 

It is also possible that the millilens could be a point mass, such as a black hole. A population of wandering black holes in cluster environments is predicted by simulations \citep{ricarte21a,ricarte21b} and is capable to producing asymmetries in magnification and missing counterimages of sources \citep{mahler23}, akin to those seen in the case of Mothra. However, the precise abundance and mass function of the wandering population are poorly constrained currently. Recent limits at $z \sim 2$ from the {\tt ASTRID} simulation find that $\sim$10-100 black holes with masses ranging from $10^4 - 10^6 M_{\odot}$ are expected near the centers of massive galaxies \citep{dimatteo23}, making them a plausible candidate for the Mothra millilens.

Alternatively, the millilens may be a dark matter subhalo source, which is consistent with expectations from $\Lambda$CDM cosmology \citep{diego23b,williams24}, which predicts numerous dark matter subhalos at pc scale. Therefore, our upper limit constraints on the millilens offer a good test for dark matter constraints, which we elaborate on in Section \ref{txt:darkmatter}. Regardless of the identity of the millilens, our results seem to be in good agreement with past results and the standard cosmological paradigm.

\subsection{Implications for Dark Matter}\label{txt:darkmatter}

The consistent presence of the light unaffiliated substructures M1 and M2 in all our models make them intriguing candidates for potential dark matter substructures. Likewise the millilens structure in Mothra, if assumed to be dominated by dark matter, can offer a unique probe of the low mass end of the mass function of dark matter haloes. We emphasize that a more detailed analysis than the one presented here would be necessary to establish the reality of these features and to formalize these constraints. Here, we perform a simple analysis under the assumption that our findings hold.

Wave dark matter (also known as fuzzy dark matter) has recently had success in reconstructing anomalous flux ratios in galaxy-scale strong lenses \citep{amruth23}, and therefore offers an interesting candidate for dark matter. In the wave dark matter formulation (hereafter $\psi$DM), dark matter is described by a scalar field with the Schr\"{o}dinger-Poisson equation \citep{hui17}. The associated particles in this model are ultra-light bosons with mass $m_{\psi} \sim 10^{-22}$ eV, corresponding to astrophysical scale de Broglie wavelengths $\lambda \sim 0.1 - 1$ kpc. At this scale, fluctuations of the density distribution can form mass substructures, referred to as granules, which oscillate on very long timescales. These granules contain soliton cores arising from the balance of quantum pressure and gravity. In general, these soliton cores have a mass $M$ and a length scale $\lambda$, which scale as \citep{schive14,burkert20,amruth23}:
\begin{equation}\label{eq:FDMdebrog}
    \lambda \propto \left(1+z_d\right)^{-1/2}m_{\psi}^{-1}M^{-1/3}.
\end{equation}
As an order of magnitude calculation, if we assume M1 and M2 are single solitons with core radii equivalent to $\lambda$, we estimate $m_{\psi} = 8.61 \pm 0.15 \times 10^{-25}$ eV and $2.04 \pm 0.02 \times 10^{-24}$ eV for M1 and M2, respectively. These are lower limits on $m_{\psi}$ because it is alternatively possible to assume that M1 and M2 are instead a clustering of many solitons. In that case, the $\psi$DM substructure mass would decrease, implying a larger $m_{\psi}$. More sophisticated analysis would be required to better constrain $\lambda$ with M1 and M2.

It is worth mentioning the recent tension in $m_{\psi}$ constraints from astrophysical observations, with some results favoring $m_{\psi} \sim 10^{-22}$ eV \citep{schive14,schive16,amruth23,diego23b} while others favor $m_{\psi} \gtrsim 10^{-21.5}$ eV \citep{irsic17,davies20,laroche22,powell23}. Therefore, a more useful constraint is an upper limit on $m_{\psi}$. We can extend our previous calculation to an upper limit on $m_{\psi}$ using results from our Mothra millilens. The subhalo mass function for $\psi$DM becomes increasingly suppressed at lower halo masses until it reaches a lower mass limit $M_{\rm min}$ due to quantum pressure \citep{hui17}. At this limit, a corresponding upper limit on $m_{\psi}$ can be solved for \citep{schive14,laroche22}:
\begin{equation}\label{eq:masslimFDM}
    m_{\psi} < 10^{-22} \left(\frac{M_{\rm min}}{1.2 \times 10^8}\right)^{-2/3} (1+z_d)^{1/2}, 
\end{equation}
where $m_{\psi}$ has units of eV and $M_{\rm min}$ is expressed in $M_{\odot}$. In Section \ref{txt:millilensmodels}, we model the Mothra millilens as a perturbation on the local density profile. As we discuss, there is expected to be a lower mass limit on the millilens at the point where the perturbation fails to produce critical curves. As our results show in Section \ref{txt:millilensresults}, $\delta_i$ appears to converge at 0.01" from Mothra, for which we found the millilens to have a mass of $1.89^{+0.02}_{-0.01} \times 10^5 M_{\odot}$. Since we expect this mass to lie near the lower limit of the millilens mass, we treat it as the rough millilens lower limit, $M_{\rm min}$, for just this calculation. This is a fair assumption because if we treat the millilens as a density fluctuation, then its amplitude is approximately equal to the  mean local density (i.e. fractional overdensity $d\rho / \rho \sim 1$), as expected from $\psi$DM \citep{dalal21}.  From this, we find that $m_{\psi} < 8.74^{+0.03}_{-0.06} \times 10^{-21}$ eV. Going one step further, simulations show that $\sim$80\% (possibly even higher since Mothra is close to the cluster center) of the mass of dark matter halos can get stripped during their infall \citep{niemiec19}, something expected to occur in the case of the Mothra millilens substructure. Accounting for this, the original mass of the millilens increases to $9.43^{0.10}_{0.05} \times 10^5 M_{\odot}$, thus constraining $m_{\psi} < 2.99^{+0.01}_{-0.02} \times 10^{-21}$ eV. 

We caution that this conclusion relies on many assumptions, mainly that our analysis adequately estimates $M_{\rm min}$, that T-strip is a good representation of the millilens, and that Mothra is indeed a binary pair of supergiants (thus requiring $\mu_m \sim 6000$). The first two of these were justified previously, while the last is our most sensitive assumption. Degeneracies present in the SED fitting of Mothra's photometry potentially cause $T_{\rm eff}$ to be underestimated \citep{diego23b}. Since our constraints of the millilens properties heavily rely on the intrinsic properties of Mothra, we would ideally prefer spectroscopic observations to characterize Mothra more accurately. However, acquisition of a spectra of Mothra is likely to pose a challenge, due to an expected long exposure time of over 50 hours per grating with JWST and lack of identifiable spectral lines at its redshift \citep{lundqvist24}. Therefore, our constraints presented here are based on the best available data, and may drastically change upon future observations. Nonetheless, the upper limit is consistent with both Lyman-$\alpha$ measurements \citep{irsic17} and lensing flux ratios \citep{amruth23}, thus unable to break the tension. Since this only makes use of a single low mass millilens substructure, the discovery of more Mothra-like millilenses \citep[e.g. Godzilla, in the Sunburst arc][]{diego22b} may help elucidate the $m_{\psi}$ upper limit to higher precision.

Another dark matter candidate worth investigating is self-interacting dark matter (SIDM). This model assumes the dark matter particles have a non-zero interaction cross-section per unit mass $\sigma/m$, which has been constrained using dark matter substructures with cosmological N-body simulations \citep{peter13,harvey19,xu23,sabarish24} and lensing observations \citep{miraldaescude02,markevitch04,bradac08,jauzac16,jauzac18,andrade22,kong24}. Once again, a more proper analysis of SIDM constraints derived from our lens model is beyond the scope of this paper. Here, we present a brief back-of-envelope calculation to constrain $\sigma/m$.

The characteristic radius of the SIDM density profile is defined as the point where dark matter self-scattering happens once during the age of the halo ($t_{\rm age}$) \citep{kaplinghat16}. If we assume that the SIDM halo is virialized at this characteristic radius and that the characteristic radius is roughly equivalent to the core radius $R_c$, then we obtain the following relation \citep{miraldaescude02,peter13,kaplinghat16}:
\begin{equation}\label{eq:dmcrossSIDM}
    \frac{\sigma}{m} = \frac{4\pi}{3t_{\rm age}}\sqrt{\frac{R_c^7}{GM^3}},
\end{equation}
where $M$ is the mass of the SIDM halo. This relation also assumes that the density of the SIDM halo at $R_c$ is equivalent to its mean density, something which is approximately true for our purposes. For this calculation we set $t_{\rm age} \approx 1.5 \times 10^9$ years which corresponds to the formation of the halo at $z \sim 0.6$. This choice of redshift is rather arbitrarily chosen such that the halo is relatively long lived, and therefore gravitationally bound, by the time of observation at $z = 0.396$. With this, we find $\sigma/m$ to be $0.070 \pm 0.005$ cm$^2$ g$^{-1}$, $0.013 \pm 0.001$ cm$^2$ g$^{-1}$, and $2.301 \pm 0.027 \times 10^{-3}$ cm$^2$ g$^{-1}$ for M1, M2, and the Mothra millilens, respectively. We note that for this calculation we use the best fitting mass and core radius for the millilens. Since these are tidally stripped substructures with original masses that are approximately estimated to be much higher, these constraints are rough upper limits. These results are stricter but nonetheless consistent with previous constraints \citep{peter13,kaplinghat16,andrade22}. It is also in agreement with recent results finding best fitting $\sigma/m \ll 1$ cm$^2$ g$^{-1}$ \citep{harvey19,andrade22}, suggesting that dark matter in clusters does not self-interact. Similar to $\psi$DM, a more rigorous and sophisticated analysis is necessary to further constrain the interaction cross section.

\section{Conclusions}\label{txt:conclusions}

We present an updated lens reconstruction (dubbed FF00) of MACSJ0416 using {\tt GRALE}, a free-form genetic algorithm based lens inversion method. FF00 includes all 237 spectroscopically confirmed images, making this the free-form lens model with the most multiple image constraints in the Frontier Fields. Our main results are as follows:

\begin{itemize}
    \item Our main lens model, FF00, has $\Delta_{RMS} = 0.191"$, indicating high accuracy in the reconstruction. We also find $M(< 200\,{\rm kpc})$ to be $1.434 \pm 0.002 \times 10^{14} M_{\odot}$ and $1.487 \pm 0.002 \times 10^{14} M_{\odot}$ for BCG-N and BCG-S, respectively, which is consistent with previous models \citep{diego23a}. Table \ref{tab:models} presents the main results of the additional models we generate. All the models are broadly consistent with one another. In addition, all the models exhibit the same large scale density structure.
    \item We identify two main dark substructures that appear to be unassociated with light (M1 and M2) that are present in all the models we generate. M1 is located roughly in the center of the lens with $9.5 \pm 0.5 \times 10^{11} M_{\odot}$ within a core radius of $\sim$16 kpc. M2 is located much closer to BCG-S with a size of $\sim$8 kpc and mass of $5.7 \pm 0.2 \times 10^{11} M_{\odot}$. While it is possible that the mass of both substructures can be redistributed using lensing degeneracies (thereby reducing or eliminating them), we argue that this is unlikely (especially for M2) due to the presence in both cases of nearby central maxima images. If the two substructures are real, the axion mass lower limit of wave dark matter for M1 and M2 is $\sim 10^{-24}$ eV. Likewise, dark matter self-interacting cross sections above $0.070 \pm 0.005$ cm$^2$ g$^{-1}$ and $0.013 \pm 0.001$ cm$^2$ g$^{-1}$, are disfavoured. In the future, we suggest more sophisticated analysis to formalize these constraints.
    \item For the well studied Spock Arc, we find FF00 exhibits two crossings of the arc, one of the only models with this behavior. The mean image plane separation $\langle \Delta \boldsymbol{\theta} \rangle$ for this arc is 0.111". There is strong support for this being an accurate reconstruction of the critical curve structure near the Spock arc due to the calculated high magnification ($\mu > 40$) across the arc. This allows for a high probability of microlensing to explain the numerous transients \citep{yan23} discovered in the arc. The multiple crossings are able to provide sufficient mass to account for the $M/L$ ratios of Spock-N and Spock-S, which imply that the wide density profile around Spock-N is responsible for the 2 critical curve crossings. Our hybrid (H-prefixed models) and transient counterimage (FF-prefixed except for FF00) models are unable to reproduce as accurate a model nor the necessary high magnification across the arc to explain the number of transients. Predicted luminosities of transients with this model are consistent with blue supergiants at the upper limit, consistent with observations \citep{rodney18,diego24}.
    \item In the Mothra arc, the presence of Mothra for $>8$ years without a counterimage suggests a unique case of millilensing at that position. Using Bayesian optimization to model this millilens, we constrain the mass and core radius of the millilens to be $2.29^{+0.91}_{-0.71} \times 10^6 M_{\odot}$ and $17.1^{+2.8}_{-2.3}$ pc at the upper limit, respectively. This result is consistent with the model from \cite{diego23b}, and inconsistent with the model from \cite{abe23}. Our result provides the first explicit constraint on the core radius of the millilens. The upper limits on mass and core radius are consistent with the millilens source being a small globular cluster too dim to be seen with JWST, or an intermediate mass black hole.  Alternatively, if the millilens is dominated by wave dark matter, we constrain the axion mass $m_{\psi} < 2.99^{+0.01}_{-0.02} \times 10^{-21}$ eV, consistent with past results \citep{irsic17,amruth23}. Similarly an upper limit of $2.301 \pm 0.027 \times 10^{-3}$ cm$^2$ g$^{-1}$ can be placed on the dark matter self-interacting cross section. Like M1 and M2, more rigorous analysis is required to formalize these constraints.
\end{itemize}

The natural next step for this work would be to do another {\tt GRALE} lens reconstruction of MACSJ0416 including the full JWST image catalog of 343 multiple images \citep{diego23a}. As mentioned in Section \ref{txt:data}, we did not include this larger image dataset because it lacks spectroscopic redshifts for many sources, which can reduce $\Delta_{\rm RMS}$ in {\tt GRALE} \citep{remolina18}. It is likely, however, that spectroscopic redshifts will be available for the larger image set in the future. Furthermore, since MACSJ0416 has the largest number of multiple images of any known cluster lens, it can serve as an excellent test on whether or not lens models in the literature are converging to a similar solution. We propose a comparison of reconstructed density profiles for this lens for various lens inversion methods and image constraints. Presumably, the increase in number of image constraints (especially in the central regions of the cluster) and precision of $M/L$ measurements for cluster member galaxies has caused mass reconstructions to converge for a variety of parametric and non-parametric lens inversions. If this is true, this can allow us to establish the likelihood of certain mass features in the lens (such as M1 or M2) being real structures if they are consistently reproduced with increasing data constraints. In turn this can help place stronger constraints on dark matter models.

In addition, the inclusion of highly magnified transients (such as with Mothra in this work) as explicit constraints on substructure in the lens is a new and exciting frontier for future research \citep{venumadhav17,dai21a,griffiths21,diego22a,lin22,meena22,williams24}. The technique can be used to place constraints on microlensing probability and density \citep{diego24} and consequently on different models of dark matter \citep{dai20b}. The multiple critical curve crossings of the Spock arc we find in FF00 offers a unique case deserving of future study. Since our model predicts a high magnification along the arc, the probability of microlensing increases. We suggest future work using our model to therefore formalize the constraint on microlensing probability and density in the arc. Likewise, the increasing number of transients discovered in the Spock arc with HST's Flashlights and JWST's PEARLS will elucidate the exact position of the critical curve, thus offering a test of the reality of our model. 

Our development of hybrid lens models using {\tt GRALE} offers a unique way  to perform a lens reconstruction with a parametric basis set constraint included on top of a free-form inversion method, which uses thousands of free parameters and generates an ensemble of solutions, instead of just one\footnote{A hybrid lens inversion method using a much smaller number of parameters and yielding a single solution is WSLAP \citep{diego05,sendra14,diego16}.}.
In our work, these models were broadly consistent with the free-form ones. In SDSS J1004+4112, \cite{perera24} found, using hybrid {\tt GRALE} models, that a denser central mass distribution offered a competing degenerate model with similar fitness, showing how hybrid models can uniquely examine the parameter space. However, the performance of the hybrid {\tt GRALE} method (and if it contains any systematic effects) has not yet been studied in detail. We therefore suggest rigorous scrutiny  of the performance of hybrid {\tt GRALE} models and how they compare with the default free-form {\tt GRALE} models in a manner similar to \cite{gho20}. Such an analysis will allow clearer interpretations of future {\tt GRALE} models which will be crucial with the discovery of greater numbers of multiple images per cluster.

In general, our lens reconstructions achieve one of the highest precision models of MACSJ0416 to date. The reconstruction of multiple critical curve crossings in the Spock arc will, if true, place strong constraints on the local density of microlenses and dark matter. Our constraints on the mass and size of the Mothra millilens likewise place constraints on models of dark matter. 
%%%%%%%%%%%%%%%%%%%%%%%%%%%%%%%%%%%%%%%%%%%%%%%%%%

\section*{Acknowledgements}
The authors acknowledge the computational resources provided by the Minnesota Supercomputing Institute (MSI), which were critical for this study. We thank Pietro Bergamini for providing necessary measurements used in this work for effective comparison with their model. The authors also would like to thank Lindsey Gordon, John Hamilton Miller Jr., and Joseph Allingham for useful discussions and suggestions regarding this work. DP acknowledges the School of Physics and Astronomy, University of Minnesota for partially supporting this work through the Robert O. Pepin Fellowship. PN is partially supported by the Black Hole Initiative at Harvard University, which is funded by the Gordon and Betty Moore Foundation grant 8273, and the John Templeton Foundation grant 61497. ML acknowledges the Centre National de la Recherche Scientifique (CNRS) and the
Centre National des Etudes Spatiale (CNES) for support.
%%%%%%%%%%%%%%%%%%%%%%%%%%%%%%%%%%%%%%%%%%%%%%%%%%

\section*{Data Availability}

Data generated from this article will be shared upon reasonable request to the corresponding author.

\bibliographystyle{mnras}
\bibliography{references}

%%%%%%%%%%%%%%%%%%%%%%%%%%%%%%%%%%%%%%%%%%%%%%%%%%

%%%%%%%%%%%%%%%%%%%%%%%%%%%%%%%%%%%%%%%%%%%%%%%%%%

%%%%%%%%%%%%%%%%% APPENDICES %%%%%%%%%%%%%%%%%%%%%

\appendix

\section{Hybrid Models}\label{txt:hybrids}
Hybrid lens models using {\tt GRALE} amount to including additional basis functions on top of the currently existing grid of Plummers. All these functions are optimized in the genetic algorithm. The benefit of these models is the mandatory inclusion of mass at the scale of individual cluster member galaxies, which can have influence on image positions and magnifications that may not be completely reconstructed by a free-form cluster scale model. Below we describe the input parameters and their physical justifications for the Sersic and NFW hybrid models.

\subsection{Sersic Model}\label{txt:sersic}
For a circular Sersic profile, the projected surface mass density \citep{keeton01} is:
\begin{equation}\label{eq:sersic}
    \Sigma_S(\boldsymbol{\theta}) = \Sigma_{\rm cen} \exp \left[ -\left(\frac{\boldsymbol{\theta}}{\theta_S}\right)^{\frac{1}{n}}\right],
\end{equation}
where $\Sigma_{\rm cen}$ is the central surface mass density, $n$ is the Sersic index, $\theta$ is the angular position, and $\theta_S$ is the angular scale, defined by the effective radius $R_e$ as $\theta_S = (b^{-n}R_e)/D_d$, where $b = 2n - (1/3)$. To model cluster member galaxies with a Sersic profile, we define for each modelled galaxy $\Sigma_{\rm cen}$, $R_e$, and $n$ based on observed physical characteristics from each galaxy. We note that the parameter $\Sigma_{\rm cen}$ need only be an estimate since the genetic algorithm will optimize its weight.

For all modelled galaxies we assume $n = 4$ for a de Vaucouleurs model since the cluster member galaxies are most likely elliptical. For MACSJ0416, \cite{tortorelli23} measure structural parameters for the cluster member galaxies. We adopt their measurements of $R_e$ in HST F160W, since elliptical galaxies' flux is brighter in redder wavelengths, so $R_e$ measurements in this band most likely correspond to stellar mass. Estimating the central density is slightly more involved as mass measurements of the cluster member galaxies are unavailable. We start by solving for $\Sigma_{\rm cen}$ in terms of the total stellar mass $M_{\star}$ (with $n = 4$):
\begin{equation}
    M_{\star} = 2\pi \Sigma_{\rm cen} D_d^2 \int \boldsymbol{\theta} \exp \left[ -\left(\frac{\boldsymbol{\theta}}{\theta_S}\right)^{\frac{1}{4}}\right] d\boldsymbol{\theta}.
\end{equation}
Since we have assumed a circular Sersic profile, we can rewrite the equation:
\begin{equation}
     M_{\star} = 2\pi\Sigma_{\rm cen} \int_0^{\infty} r \exp \left[ -\left(\frac{r}{R_S}\right)^{\frac{1}{4}}\right] dr,
\end{equation}
where $r$ is the radial distance with respect to the center of the galaxy ($r = D_d\theta)$ and $R_S$ is the physical scale, defined as $R_S = D_d\theta_S = b^{-4}R_e$. The solution for this is exact if one substitutes $x = (r/R_S)^{1/4}$:
\begin{align*}
    M_{\star} &= 2\pi\Sigma_{\rm cen} \int_0^{\infty} x^4 R_S e^{-x} \left(4x^3 R_S \right) dx \\
    &= 8\pi\Sigma_{\rm cen} R_S^2 \int_0^{\infty} x^7 e^{-x} dx \\
    &= 8!\pi\Sigma_{\rm cen}R_S^2.
\end{align*}
So to estimate $\Sigma_{\rm cen}$ we need to estimate the stellar mass:
\begin{equation}
    \Sigma_{\rm cen} = \frac{M_{\star}}{8!\pi R_S^2}.
\end{equation}
Observations from the Canada-France-Hawaii Telescope Legacy Survey (CFHTLS) of cluster member galaxies at $0.1 < z < 0.9$ find a tight correlation between $R_e$ and $M_{\star}$ \citep{ulgen22}. Using our adopted estimates for $R_e$, we can therefore use this relation \citep[see equation 9 in][]{ulgen22} to estimate $M_{\star}$ and therefore $\Sigma_{\rm cen}$. Table \ref{tab:hybridS} gives the estimated Sersic parameters used as input for our hybrid models.

\subsection{NFW Model}\label{txt:nfw}
The Navarro-Frenk-White (NFW) profile describes the dark matter density profile for a dark matter halo \citep{nfw97}:
\begin{equation}\label{eq:NFW3ddens}
    \rho_{\rm NFW}(r) = \frac{\rho_s}{\frac{r}{r_s}\left(1 + \frac{r}{r_s}\right)^2},
\end{equation}
where $r_s$ is the scale radius and $\rho_s$ is the scale density. The projected surface density for an NFW profile is \citep{bartelmann96,nfw97,wright00}:
\begin{equation}\label{eq:nfw}
    \Sigma_{\rm NFW}(\theta) = 2r_s\rho_s G\left(\frac{\theta}{\theta_s}\right),
\end{equation}
where $\theta_s = r_s/D_d$ and $G(x)$ is defined as:
\begin{equation}
    G(x) = \frac{1 - F(x)}{x^2 - 1}
\end{equation}
 with
\begin{equation}
    F(x) = \left\{ \begin{array}{lrc} \frac{1}{\sqrt{1-x^2}} \tanh^{-1}{\sqrt{1-x^2}} & \mbox{for} & x<1 \\ 1 & \mbox{for} & x = 1 \\
    \frac{1}{\sqrt{x^2-1}} \tan^{-1}{\sqrt{x^2-1}} & \mbox{for} & x>1.
\end{array}\right.
\end{equation}
From this we need to estimate $r_s$ and $\rho_s$ in order to model the cluster member galaxies as NFW profiles. 

For a physically realistic model, we choose to define the scale radius as the virial radius. From this assumption, we can solve for the virial mass of the NFW profile:
\begin{align*}
    M_{\rm vir} &= \int_0^{R_{\rm vir}} 4\pi r^2 \rho_{\rm NFW}(r) dr \\
    &= 4\pi \rho_s r_s^3 \left(\ln\left(1+c\right) - \frac{1}{1+c}\right),
\end{align*}
where $c$ is the concentration parameter defined as $c = R_{\rm vir}/r_s$. To get the virial radius, we estimate it using the effective radius and virial radius relation from \cite{huang17} where effective radius $R_e \sim 0.023 R_{\rm vir}$. Using the measured $R_e$ of the cluster member galaxies from \cite{tortorelli23}, and this relation, we can estimate the scale radii for the cluster member galaxies:
\begin{equation}
    r_s \approx \frac{R_e}{0.023c}. 
\end{equation}
From this, the scale radius for a given galaxy depends on its respective concentration parameter, which we estimate using the concentration-mass relation as measured in \cite{correa15}. This requires an estimate of the virial mass independent of that of the NFW profile such that $c$ is physically justified rather than an assumption. Since, from the virial theorem, the virial mass can be estimated as:
\begin{equation}\label{eq:virial}
    M_{\rm vir} = \frac{5R_{\rm vir}\sigma^2}{G},
\end{equation}
an estimate of the velocity dispersion $\sigma$ will give us $c$. We note that equation \ref{eq:virial} is a good approximation for elliptical galaxies \citep{cappellari06,bezanson15} Velocity dispersion is not directly measured for some of the galaxies in MACSJ0416, so we estimate it by scaling these with the measured velocity dispersion of BCG-N $\sigma_0 = 279$ km s$^{-1}$ \citep{bergamini21} and the best fit Faber-Jackson relation for MACSJ0416 \citep{bergamini21}. Thus, for a given galaxy, we have the following relation (from Faber-Jackson):
\begin{equation}
    \frac{L_{\rm gal}}{L_0} = 10^{-0.4(M_{\rm gal}-M_0)} = \left(\frac{\sigma}{\sigma_0}\right)^{\frac{1}{\alpha}},
\end{equation}
where $M_{\rm gal}$ and $M_0$ are the absolute magnitudes for a chosen galaxy and BCG-N, respectively, $L_{\rm gal}$ and $L_0$ are the luminosities for a chosen galaxy and BCG-N, respectively, and $\alpha = 0.3$ \citep{bergamini21} for the best fitting Faber-Jackson relation for the cluster. $M_0$ can be solved for with the distance modulus using the HST F160W photometric measurement of the apparent magnitude of BCG-N $m_0 = 17.02$. Similarly, photometric apparent magnitudes in the F160W band for each cluster member galaxy are presented in \cite{tortorelli23}, allowing for measurements of $M_{\rm gal}$. From this, we can get the velocity dispersion for a given cluster member galaxy:
\begin{equation}\label{eq:veldisp}
    \sigma = \sigma_0 \left(10^{-0.4\alpha(M_{\rm gal}-M_0)}\right).
\end{equation}

To summarize, using photometric measurements in HST F160W, we are able to obtain estimates of each cluster member galaxy's velocity dispersion with equation \ref{eq:veldisp}. From this, we can estimate the virial mass using equation \ref{eq:virial}. Lastly, we can use this estimate in the concentration-mass relation from \cite{correa15} to estimate $c$. At this point, we are able to then obtain the scale radius $r_s$. The final step to model the NFW is to solve for $\rho_s$, which is done simply by equating the virial mass from the integrated NFW profile with our estimate from virial theorem (equation \ref{eq:virial}):
\begin{align*}
    \frac{5R_{\rm vir}\sigma^2}{G} &= 4\pi \rho_s r_s^3 \left(\ln\left(1+c\right) - \frac{1}{1+c}\right).
\end{align*}
Thus, we obtain:
\begin{equation}
    \rho_s = \frac{5R_{\rm vir}\sigma^2}{4\pi r_s^3 G} \left(\ln\left(1+c\right) - \frac{1}{1+c}\right)^{-1}.
\end{equation}
Table \ref{tab:hybridN} gives the estimated NFW parameters used as input for our hybrid models.

\section{Uncertainty in the Spock Arc Critical Curve}\label{txt:spockunc}

\begin{figure}
\includegraphics[trim={5.6cm 0.35cm 5.2cm 0.35cm},clip,width=0.49\textwidth]{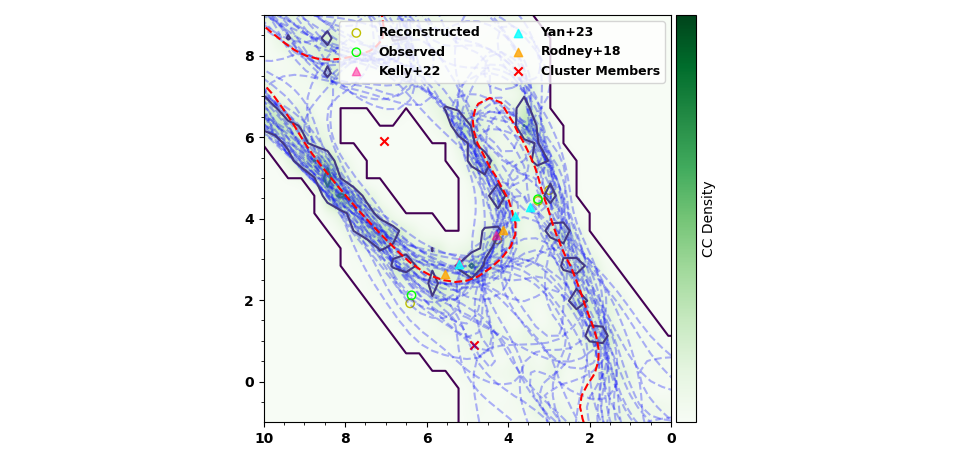}

\caption{The same field of view of the Spock arc as Figure \ref{fig:spockCC}, with the FF00 critical curve (CC) shown as a red dashed line. The labelled points are the same as in Figure \ref{fig:spockCC}. The green background (and corresponding dark purple contour lines) shows the bootstrapped density of CCs from the ensemble of 40 {\tt GRALE} runs, with darker green regions corresponding to regions where individual CCs from a {\tt GRALE} run are more likely to pass through. The blue dashed lines show the resampled CCs. From this, we can see that the FF00 CC falls along a probable path around the Spock arc. The CC density also favors multiple crossings of the arc as shown by the CC density forming a ``U'' shaped structure as it passes the arc.} 
\label{fig:spockunc}
\end{figure}

Since the critical curve structure around the Spock arc is of particular interest, it is important to analyze the uncertainty in our recovered model. Our FF00 reconstruction finds multiple critical curve crossings of the Spock arc. We bootstrap by randomly sampling the ensemble of 40 {\tt GRALE} runs with replacement,  to obtain the frequency at which multiple crossings are recovered by {\tt GRALE}. From this resampled data, we calculate the density of critical curves around the Spock arc, as shown in Figure \ref{fig:spockunc}. The critical curve density is thus a measure of the uncertainty. 

The critical curve density follows our FF00 result, and does seem to favor multiple crossings of the Spock arc. This can be seen with the valley-like features of the critical curve density. Unlike the single critical curve crossing case of \cite{bergamini23} (see top panel of Figure \ref{fig:spockCC}) and several other models, whose critical curve closely encircles the Spock-N galaxy, our critical curve density is very low around Spock-N. Our critical curve density instead increases significantly around the valley surrounding Spock-N. We can see this forms two distinct crossing points on either side of the Spock arc. We note that this does not rule out a single crossing model, as a little under half of the models in the ensemble of 40 {\tt GRALE} runs reconstructed one crossing. Augmenting the ensemble of models from which we construct FF00 will further elucidate the critical curve structure around the Spock arc.

\section{Hybrid and Free-Form Lens Model Results}\label{txt:hybridsextra}

Here we present summary plots for the transient counterimage and hybrid models we generated to compare with FF00. The transient counterimage models are defined in Section \ref{txt:input}. The hybrid models are likewise described in Section \ref{txt:input}, with input parameters listed in Tables \ref{tab:hybridS} and \ref{tab:hybridN} and derived in Section \ref{txt:hybrids}.

\setlength{\columnsep}{0pt}
\begin{figure*}
\begin{multicols}{3}
    \includegraphics[trim={5.1cm 0.35cm 5.1cm 0.35cm},clip,width=\linewidth]{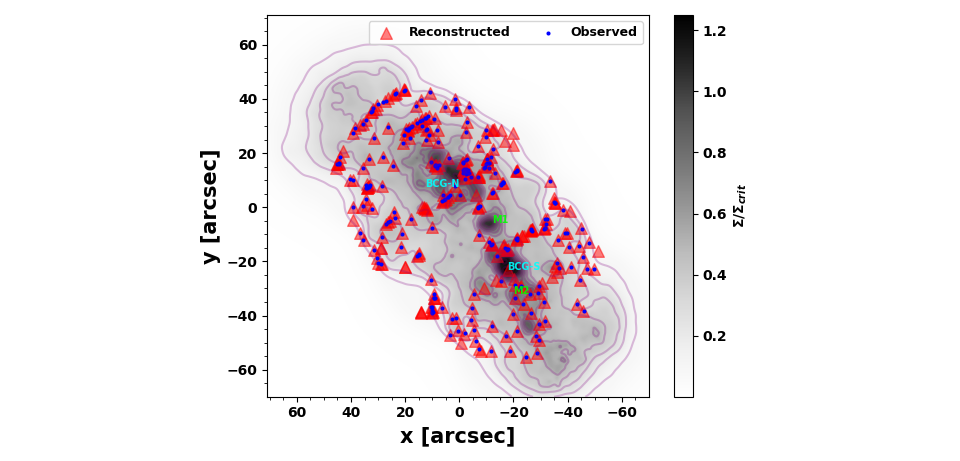}\par 
    \includegraphics[trim={4.2cm 0.42cm 4.5cm 0.6cm},clip,width=\linewidth]{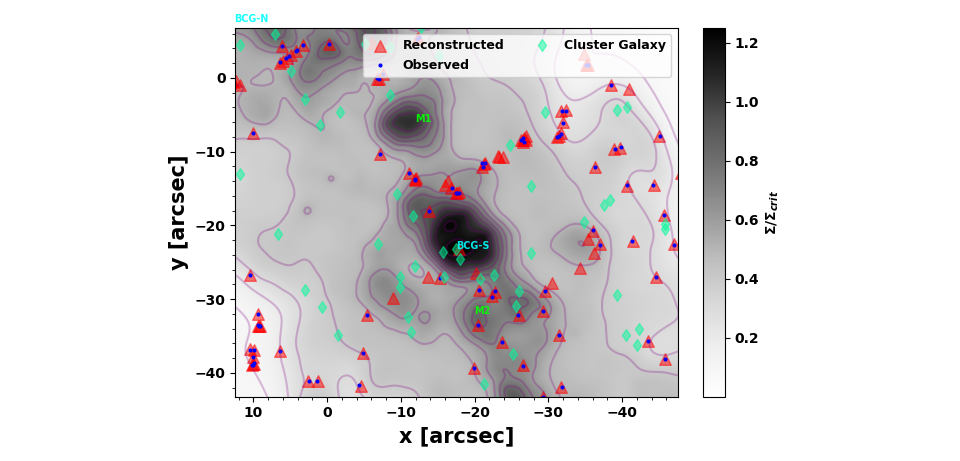}\par 
    \includegraphics[trim={5.6cm 0.35cm 5.2cm 0.35cm},clip,width=\linewidth]{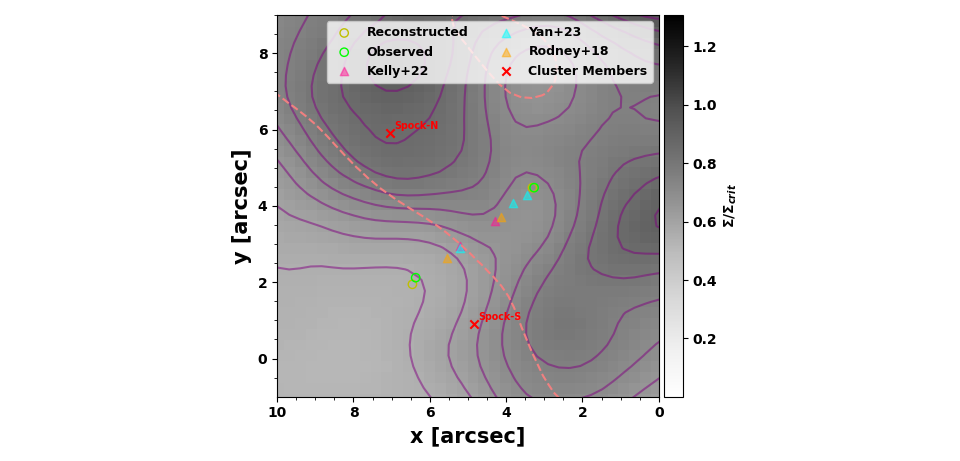}\par 
    \end{multicols}
\begin{multicols}{3}
    \includegraphics[trim={5.1cm 0.35cm 5.1cm 0.35cm},clip,width=\linewidth]{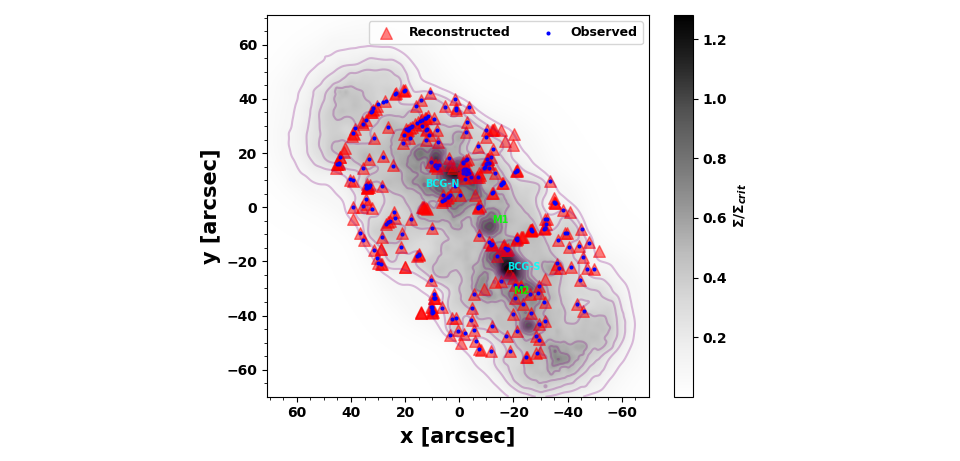}\par 
    \includegraphics[trim={4.2cm 0.42cm 4.5cm 0.6cm},clip,width=\linewidth]{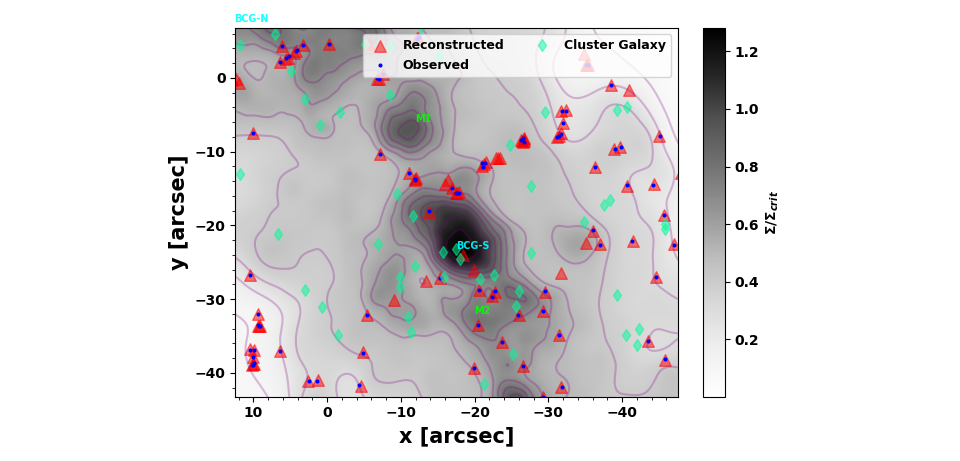}\par 
    \includegraphics[trim={5.6cm 0.35cm 5.2cm 0.35cm},clip,width=\linewidth]{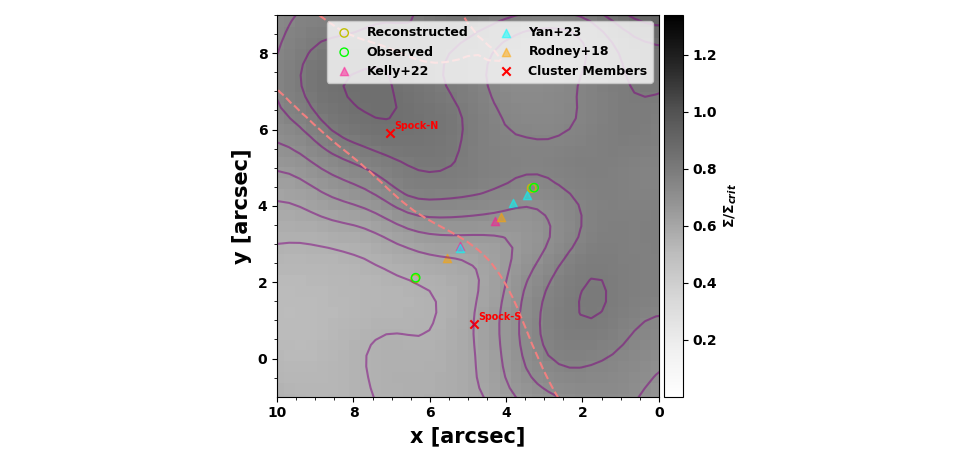}\par 
    \end{multicols}
\begin{multicols}{3}
    \includegraphics[trim={5.1cm 0.35cm 5.1cm 0.35cm},clip,width=\linewidth]{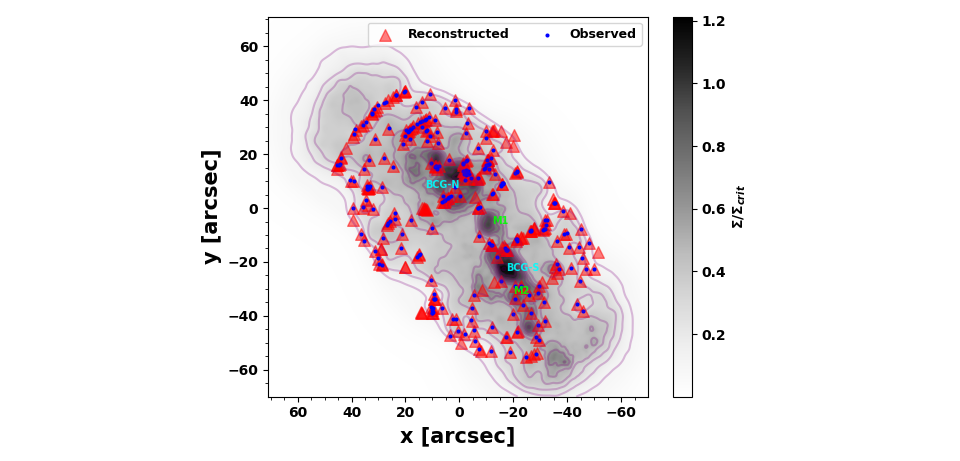}\par 
    \includegraphics[trim={4.2cm 0.42cm 4.5cm 0.6cm},clip,width=\linewidth]{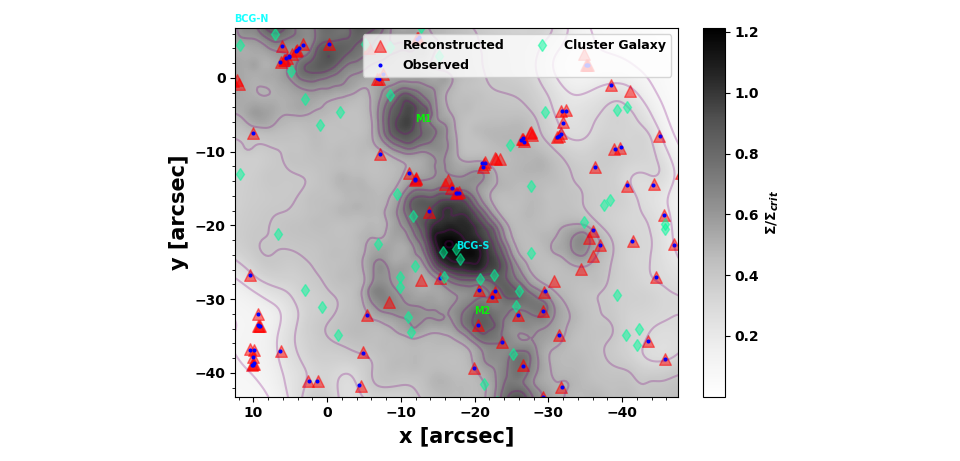}\par 
    \includegraphics[trim={5.6cm 0.35cm 5.2cm 0.35cm},clip,width=\linewidth]{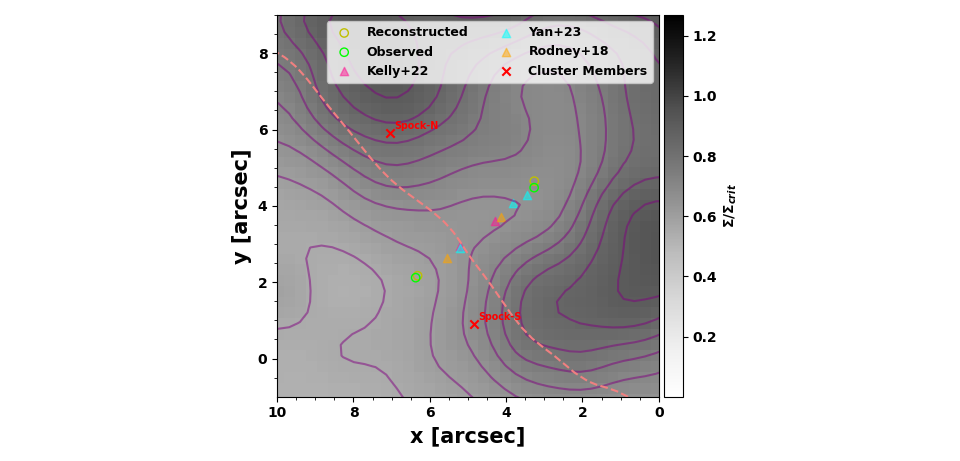}\par 
    \end{multicols}
\begin{multicols}{3}
    \includegraphics[trim={5.1cm 0.35cm 5.1cm 0.35cm},clip,width=\linewidth]{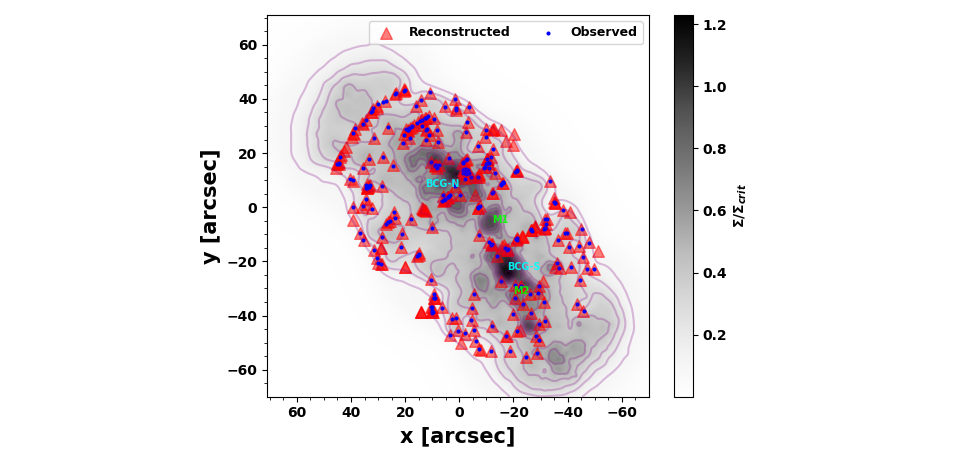}\par 
    \includegraphics[trim={4.2cm 0.42cm 4.5cm 0.6cm},clip,width=\linewidth]{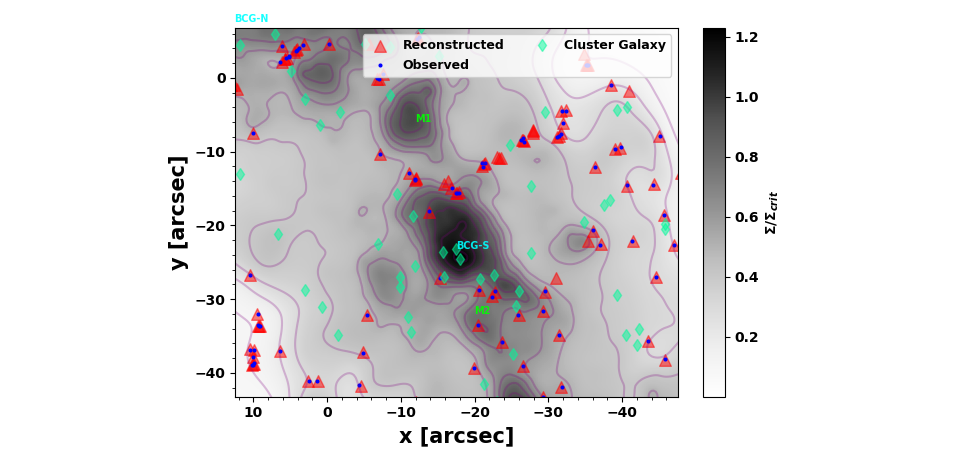}\par 
    \includegraphics[trim={5.6cm 0.35cm 5.2cm 0.35cm},clip,width=\linewidth]{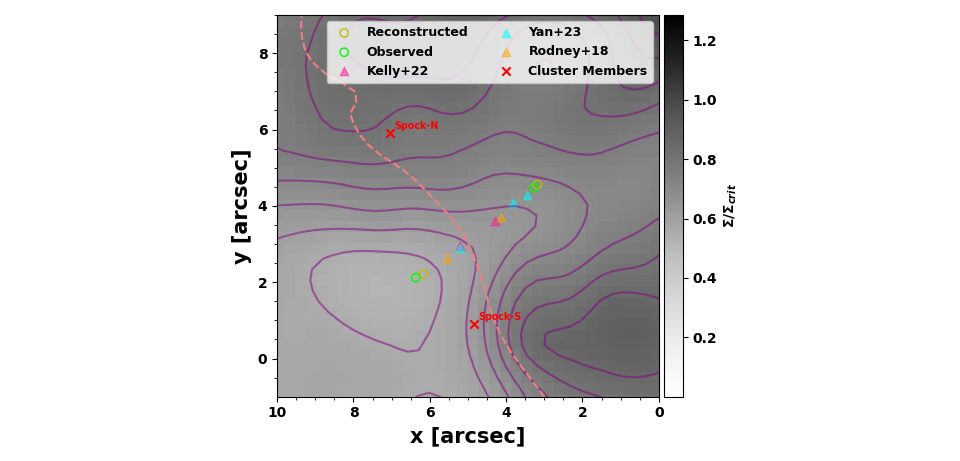}\par 
    \end{multicols}
\caption{Summary plots for the transient counterimage free-form models FF11, FF12, FF11+D, and FF12+D, from top to bottom row. ({\it Left Column:}) Projected surface mass density distribution for the respective model. Both BCGs, M1, and M2 are all labelled. Reconstructed images are shown as red triangles while observed images are shown as blue dots. ({\it Middle Column:}) Zoomed in view of the density profile in the region around BCG-S. Cluster member galaxies are shown as light green diamonds. The potential dark matter substructures M1 and M2 are labelled, and can be easily seen as light unaffiliated in this view. ({\it Right Column:}) Spock arc reconstruction for the respective model. The critical curve is shown as a dashed light red line. Observed (yellow circles) and reconstructed (green circles) images are shown alongside S1/S2 (orange triangles), F1/F2 (pink triangles), and the transients from \citet{yan23} (cyan triangles). Notable is the lack of multiple critical curve crossings of the arc contributing to low magnification along the Spock arc.}
\label{fig:FFextra}
\end{figure*}

\setlength{\columnsep}{0pt}
\begin{figure*}
\begin{multicols}{3}
    \includegraphics[trim={5.1cm 0.35cm 5.1cm 0.35cm},clip,width=\linewidth]{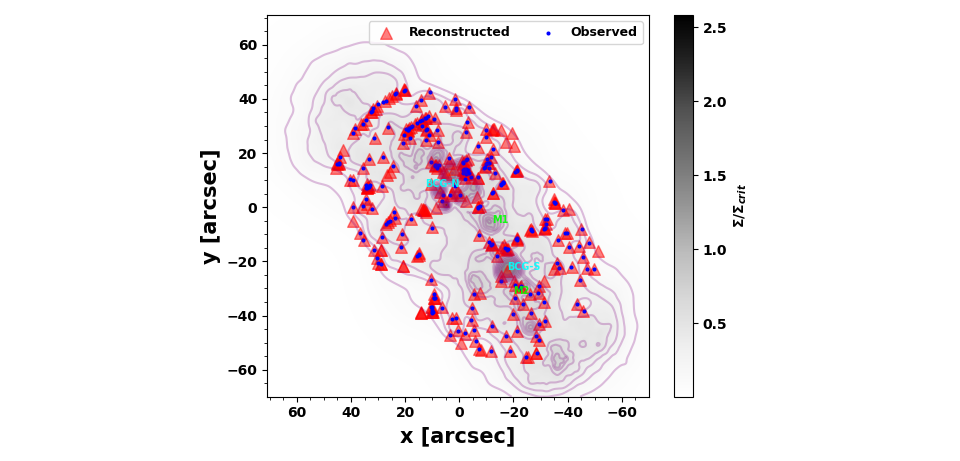}\par 
    \includegraphics[trim={4.2cm 0.42cm 4.5cm 0.6cm},clip,width=\linewidth]{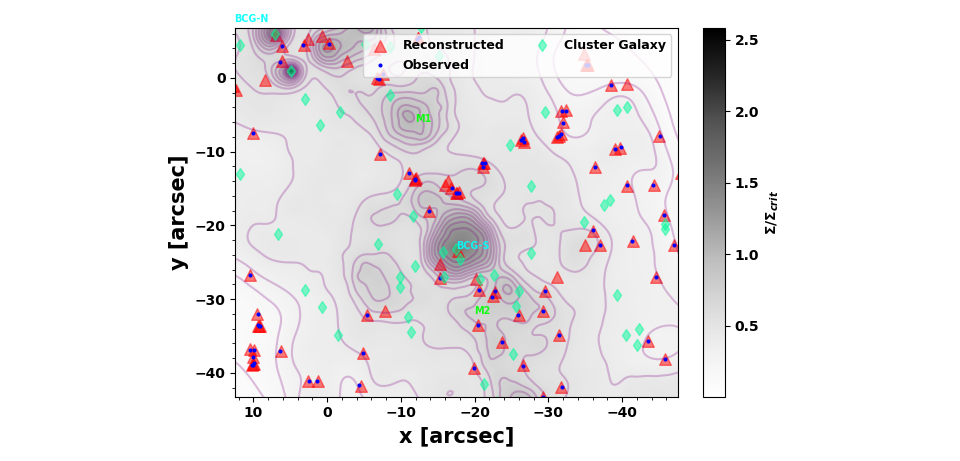}\par 
    \includegraphics[trim={5.6cm 0.35cm 5.2cm 0.35cm},clip,width=\linewidth]{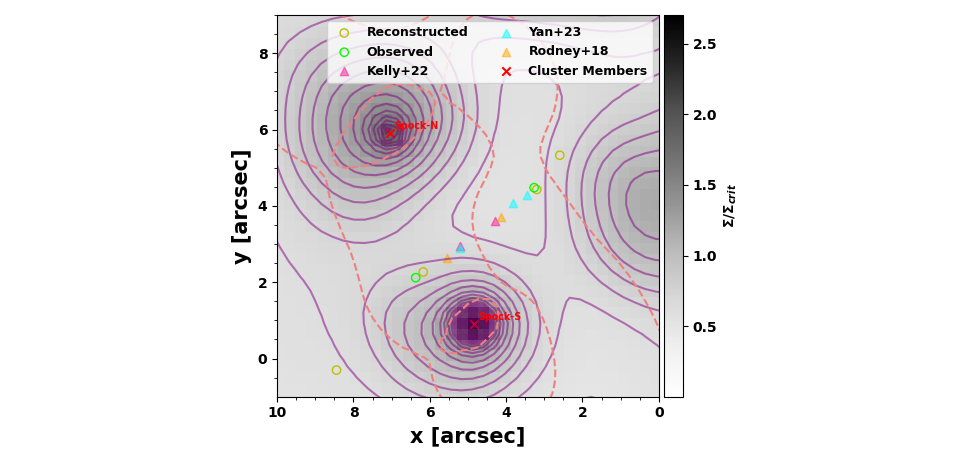}\par 
    \end{multicols}
\begin{multicols}{3}
    \includegraphics[trim={5.1cm 0.35cm 5.1cm 0.35cm},clip,width=\linewidth]{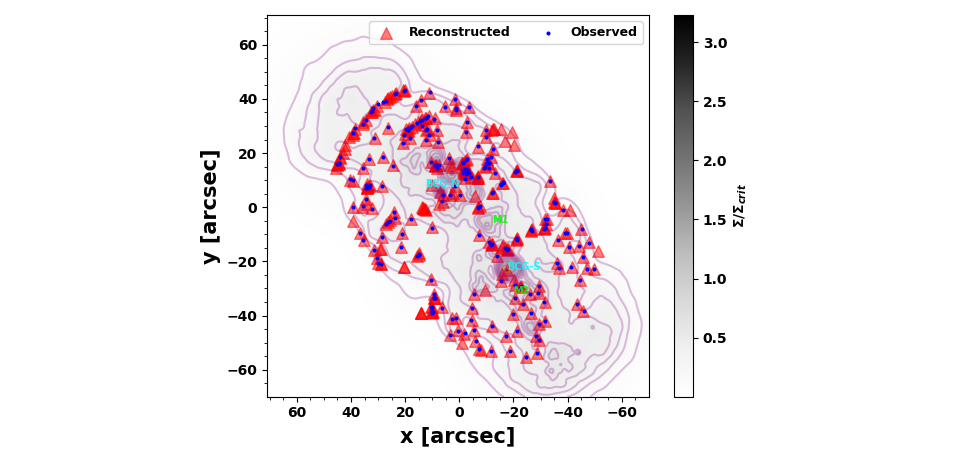}\par 
    \includegraphics[trim={4cm 0.37cm 4cm 0.1cm},clip,width=\linewidth]{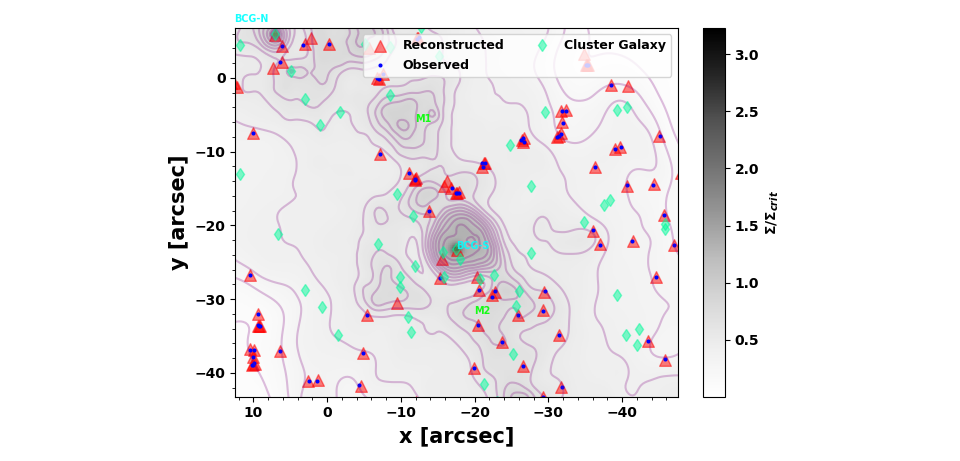}\par 
    \includegraphics[trim={5.6cm 0.35cm 5.2cm 0.35cm},clip,width=\linewidth]{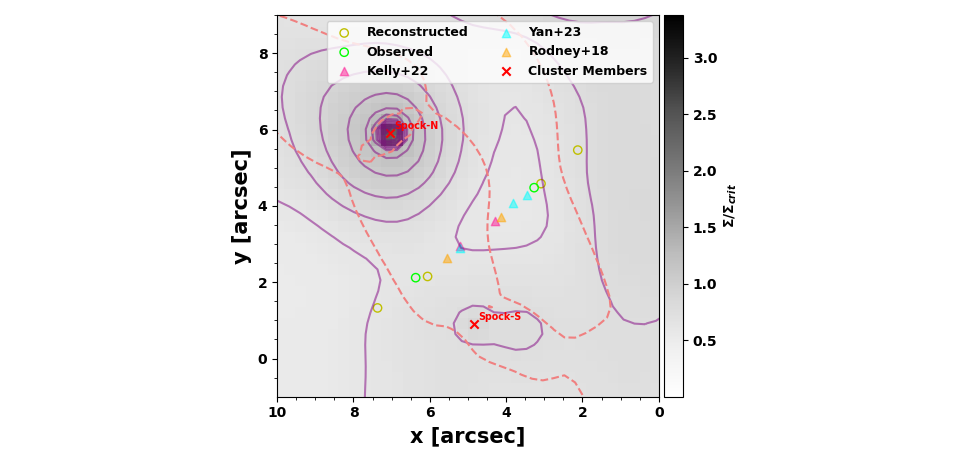}\par 
    \end{multicols}
\caption{Summary plots for the Hybrid models H-NFW ({\it Top Row}) and H-Ser ({\it Bottom Row}). Columns are the same as Figure \ref{fig:FFextra}. Note that the gray scale range representing projected density is different compared to the previous figure.}
\label{fig:Hextra}
\end{figure*}

%%%%%%%%%%%%%%%%%%%%%%%%%%%%%%%%%%%%%%%%%%%%%%%%%%

% Don't change these lines
\bsp	% typesetting comment
\label{lastpage}
\end{document}